\documentclass[useAMS,usenatbib]{mn2e}
\usepackage{times,mathptm}
\usepackage{lscape}
\usepackage{psfig}
\usepackage{subfigure}
\usepackage{multirow}

\usepackage{psfig}
\usepackage{times}
\usepackage{amsmath}
\usepackage{amssymb}
\usepackage{graphicx}
\usepackage{graphics}
\usepackage{rotating}
\usepackage{lscape}
\usepackage{morefloats}

\usepackage{pifont}
\newcommand{\cmark}{\ding{51}}%
\newcommand{\xmark}{\ding{55}}%

\def\apj{ApJ}
\def\apjl{ApJL}
\def\mnras{MNRAS}
\def\pasp{PASP}

\def\araa{ARAA}
\def\aap{A\&A}
\def\aapr{A\&AR}
\def\aj{AJ}
\def\apjs{ApJS}

\def\nat{Nature}

\def\nar{New Astron. Rev.}

\def\gs{\mathrel{\raise0.35ex\hbox{$\scriptstyle >$}\kern-0.6em
\lower0.40ex\hbox{{$\scriptstyle \sim$}}}}
\def\ls{\mathrel{\raise0.35ex\hbox{$\scriptstyle <$}\kern-0.6em
\lower0.40ex\hbox{{$\scriptstyle \sim$}}}}

\def\kms{\,\hbox{km\,s$^{-1}$}}

\def\Wm2{\,\hbox{W}\,\hbox{m}^{-2}}
\def\gsim{\mathrel{\raise0.35ex\hbox{$\scriptstyle >$}\kern-0.6em\lower0.40ex\hbox{{$\scriptstyle \sim$}}}}
\def\lsim{\mathrel{\raise0.35ex\hbox{$\scriptstyle <$}\kern-0.6em\lower0.40ex\hbox{{$\scriptstyle \sim$}}}}
\def\ltsima{$\; \buildrel < \over \sim \;$}
\def\simlt{\lower.5ex\hbox{\ltsima}}
\def\gtsima{$\; \buildrel > \over \sim \;$}
\def\simgt{\lower.5ex\hbox{\gtsima}}

\def\mum{$\mu$m}

\begin{document}

\title[KASH$z$: the prevalence of ionised outflows]{
The KMOS AGN Survey at High redshift (KASH$z$): the prevalence and
drivers of ionised outflows in the host galaxies of X-ray AGN\thanks{Based on
  observations obtained at the Very Large Telescope of the
  European Southern Observatory. Programme IDs: 086.A-0518; 087.A-0711; 088.B-0316;
  60.A-9460; 092.A-0884; 092.A-0144; 092.B-0538; 093.B-0106; 094.B-0061 and 095.B-0035.}}
\author[C.\ M.\ Harrison et al.]
{ \parbox[h]{\textwidth}{ 
C.\ M.\ Harrison,$^{\! 1\, \dagger}$
D.\ M.\ Alexander,$^{\! 1}$
J.\ R.\ Mullaney,$^{\! 2}$
J.\ P.\ Stott,$^{\! 3,1}$
A.\ M.\ Swinbank,$^{\! 4,1}$
V.\ Arumugam,$^{\! 5}$
F.\ E.\ Bauer,$^{\! 6,7,8}$
R.\ G.\ Bower,$^{\! 4,1}$
A.\ J.\ Bunker,$^{\! 3,9}$
R.\ M.\ Sharples$^{\! 10,1}$
}
\vspace*{6pt} \\
$^1${Centre for Extragalactic Astronomy, Department of Physics, Durham University, South Road,
  Durham, DH1 3LE, U.K.}\\
$^2${Department of Physics and Astronomy, University of
  Sheffield, Sheffield, S3 7RH, U.K.}\\
$^3${Astrophysics, Department of Physics, University of Oxford, Keble Road, Oxford, OX1 3RH, U.K.}\\
$^4${Institute for Computational Cosmology, Department of Physics, Durham University, South Road,
  Durham, DH1 3LE, U.K.}\\
$^5${European Southern Observatory, Karl-Schwarzschild-Strasse 2, D-85748 Garching, Germany}\\
$^6${Instituto de Astrof\'{i}sica, Facultad de F\'{i}sica,
  Pontifica Universidad Cat\'{o}lica de Chile, 306, Santiago 22,
  Chile}\\
$^7${Millennium Institute of Astrophysics, Vicu\~{n}a
  Mackenna 4860, 7820436 Macul, Santiago, Chile}\\
$^8${Space Science Institute, 4750 Walnut Street, Suite 205, Boulder,
  CO 80301, USA}\\
$^9$Affiliate Member, Kavli Institute for the Physics and Mathematics of the Universe (WPI), Todai Institutes for Advanced
Study, The University of Tokyo,\\
5-1-5 Kashiwanoha, Kashiwa, Japan 277-8583\\
$^{10}${Centre for Advanced Instrumentation, Department of Physics, Durham University, South Road,
  Durham, DH1 3LE, U.K.}\\
$^{\dagger}$Email: c.m.harrison@mail.com \\
}
\maketitle

\begin{abstract}
We present the first results from the KMOS AGN Survey at High redshift
(KASH$z$), a VLT/KMOS integral-field spectroscopic (IFS) survey of
$z\gtrsim0.6$ AGN. We present galaxy-integrated spectra of 89
X-ray AGN ($L_{{\rm 2-10keV}}=10^{42}$--10$^{45}$\,erg\,s$^{-1}$), for which
we observed [O~{\sc iii}] ($z$$\approx$1.1--1.7) or H$\alpha$ emission
($z$$\approx$0.6--1.1). The targets have X-ray luminosities representative of the parent
AGN population and we explore the emission-line luminosities as a function
of  X-ray luminosity. For the [O~{\sc iii}] targets,
$\approx$50\,per\,cent have ionised gas velocities indicative
of gas that is dominated by outflows and/or highly turbulent
  material (i.e., overall line-widths $\gtrsim$$600$\,km\,s$^{-1}$). The most
luminous half (i.e., $L_{X}>6\times10^{43}$\,erg\,s$^{-1}$) have a $\gtrsim$2
times higher incidence of such velocities. On the basis of our results, we find no
  evidence that X-ray obscured AGN are more likely to host extreme kinematics than unobscured AGN. Our KASH$z$ sample
has a distribution of gas velocities that is
consistent with a luminosity-matched sample of $z<0.4$ AGN. This implies little evolution in the prevalence of
ionised outflows, for a fixed AGN luminosity, despite an
order-of-magnitude decrease in average star-formation rates over
this redshift range. Furthermore, we compare our H$\alpha$ targets to a redshift-matched sample of star-forming
galaxies and despite a similar distribution of H$\alpha$ luminosities
and likely star-formation rates, we find extreme ionised gas
velocities are up to $\approx$10$\times$ more prevalent in the
AGN-host galaxies. Our results reveal a high prevalence of
  extreme ionised gas velocities in high-luminosity X-ray AGN and imply that
the most powerful ionised outflows in high-redshift galaxies are driven by AGN activity.
\end{abstract}

\begin{keywords}
  galaxies: active; --- galaxies: kinematics and dynamics; ---
  quasars: emission lines; --- galaxies: evolution
\end{keywords}

\section{Introduction}
Massive galaxies are now known to host supermassive black
holes (SMBHs) at their centres. These SMBHs grow through mass accretion events,
during which, they become visible as active galactic nuclei
(AGN). A variety of indirect observational
evidence has been used to imply a connection
between the growth of SMBHs and the growth of the galaxies that they reside in. For
example: (1) the cosmic evolution of volume-averaged SMBH growth and
star formation look very similar; (2) growing SMBHs may be preferentially located in
star-forming galaxies and (3) SMBH mass is tightly correlated with galaxy bulge mass and stellar velocity dispersion
(see reviews in e.g., \citealt{Alexander12};
\citealt{Kormendy13}). Suggestion of a more {\em direct} connection between SMBH
growth and galaxy growth largely comes from theoretical models of galaxy
formation. Most successful models propose that AGN are required to regulate the growth of
massive galaxies by injecting a fraction of their accretion
energy into the surrounding intergalactic medium (IGM) or interstellar
medium (ISM; e.g., \citealt{Silk98}; \citealt{Benson03}; \citealt{Granato04}; \citealt{Churazov05};
\citealt{Bower06,Bower08}; \citealt{Hopkins06};
\citealt{Somerville08}; \citealt{McCarthy10};
\citealt{Gaspari11}; \citealt{Vogelsberger14};
\citealt{Schaye15}). Without this so-called ``AGN feedback'', these models
fail to reproduce many key observables of the local Universe, such as
the observed SMBH mass-spheroid mass relationship (e.g., \citealt{Kormendy13}); the sharp
cut-off in the galaxy mass function (e.g., \citealt{Baldry12}) and the
X-ray temperature-luminosity relationship observed in galaxy clusters and
groups (e.g., \citealt{Markevitch98}).

In recent years there has been a large amount of observational work
searching for signatures of ``AGN feedback'' and to test theoretical predictions (see reviews in e.g.,
\citealt{Alexander12}; \citealt{Fabian12}; \citealt{McNamara12};
\citealt{Heckman14}). One of the most promising candidates for a
universal feedback mechanism is AGN-driven outflows, which, if they
can be driven to galaxy-wide scales, could remove or heat cold gas that would otherwise form stars in the host
galaxy. There is now wide-spread observational evidence that galaxy-wide outflows exist
in both low- and high-redshift AGN-host galaxies, using tracers of atomic, molecular and ionised
gas (e.g., \citealt{Martin05}; \citealt{Rupke05c};
\citealt{Liu13b}; \citealt{Harrison12a,Harrison14b};
\citealt{Veilleux13}; \citealt{Cicone14};
\citealt{Arribas14}). 

Of specific relevance to our study are ionised outflows that have been
known for several decades to be identifiable using broad and
asymmetric emission-line profiles (e.g., \citealt{Weedman70};
\citealt{Stockton76}; \citealt{Heckman81}). However, it is now possible to use large optical spectroscopic samples, such as the Sloan Digital Sky
Survey (SDSS; \citealt{York00}), to search for these signatures in hundreds to thousands of
$z\lesssim1$ AGN. By combining these spectroscopic surveys with multi-wavelength data sets, recent studies have provided excellent constraints on the
prevalence and drivers of ionised outflows in low-redshift AGN (e.g.,
\citealt{Mullaney13}; \citealt{Zakamska14}; \citealt{Balmaverde15}). Follow-up integral-field spectroscopy (IFS) observations of objects drawn from these large samples,
have made it possible to constrain the prevalence of these outflows on
{\em galaxy-wide} scales, to explore their spatially-resolved characteristics as a function of AGN and host-galaxy properties
and to test theoretical predictions (e.g., \citealt{Liu13b}; \citealt{Harrison14b,Harrison15}; \citealt{McElroy15}). These
studies have revealed that galaxy-wide ionised outflows are
common, perhaps ubiquitous, throughout the most optically-luminous
low-redshift AGN (i.e., with $L_{{\rm AGN}}\gtrsim10^{45}$\,erg\,s$^{-1}$; although
also see \citealt{Husemann13}).

Despite the great insight provided by statistical IFS studies of outflows
in the low-redshift Universe, these studies do not cover the redshift
ranges during the peak epochs of SMBH and galaxy growth (i.e.,
$z\gtrsim$1; e.g., \citealt{Aird10}; \citealt{Madau14}) and
consequently the redshift ranges where AGN-driven outflows are
predicted to be most prevalent. Spatially-resolved spectroscopy of high-redshift AGN has been more
limited because the bright optical emission lines, that are excellent
traces of ionised gas kinematics, are redshifted
to the near infrared (NIR), which is much more challenging to observe in than optical wavelengths. Each study that has searched for galaxy-wide ionised
outflows at high redshifts has investigated only a small number of AGN, which were selected in a
variety of different ways (e.g., \citealt{Nesvadba06,Nesvadba08}; \citealt{Alexander10};
\citealt{Harrison12a}; \citealt{CanoDiaz12};
\citealt{ForsterSchreiber14}; \citealt{Brusa15};
\citealt{Cresci15}; \citealt{Carniani15}; \citealt{Collet15}); with the largest, to date, coming from
\cite{Genzel14}, which presents 18 confirmed AGN identified using a combination of X-ray, infrared and
radio techniques. To make robust conclusions about the AGN population as a
whole, and to properly understand the role of ionised outflows in galaxy evolution, it is crucial to place
these observations into the context of the parent
population of AGN and galaxies. For example, it is particularly
important to assess how representative observations are if they have significant implications for our understanding of AGN feedback,
such as possible evidence that star formation has been suppressed by
ionised outflows in two high-redshift AGN (\citealt{CanoDiaz12};
\citealt{Cresci15}). There is clearly a need for IFS observations
of large samples of high-redshift AGN that are selected in a uniform way.

It is now possible to efficiently obtain large samples of NIR IFS data thanks to the commissioning of the K-band Multi Object Spectrograph (KMOS;
\citealt{Sharples04,Sharples13}) on the European Southern Observatory's Very Large Telescope
(VLT). This instrument is ideal for systematic studies of
the rest-frame optical properties of high-redshift galaxies and AGN,
selected in the well studied extragalactic deep fields (e.g. \citealt{Sobral13};
\citealt{Stott14}; \citealt{Wisnioski15}; Stott et~al. 2015). In this
work we present the initial results from our survey of
high-redshift AGN: the KMOS AGN Survey at High-$z$
(KASH$z$). KASH$z$ is an ongoing guaranteed time project, led by
Durham University, to observe high-redshift ($z$$\approx$0.6--3.6) AGN
with KMOS. This survey will provide a huge leap
forward for our understanding of AGN outflows and host-galaxy
kinematics, by measuring the spatially-resolved kinematics of an order
of magnitude more sources that previous work. Furthermore,
this survey has been jointly run with the KMOS Redshift One
Spectroscopic Survey (KROSS) of high-redshift star-forming galaxies
(Stott et~al. 2015), which makes it possible to place our observations
of AGN into the context of the galaxy population. 

In this paper we present the first results
from KASH$z$. In Section~\ref{sec:survey} we describe the survey and observations; in
Section~\ref{sec:analyses} we describe our data analysis and
comparison samples; in
Section~\ref{sec:results} we present our initial results and discuss their implications and in
Section~\ref{sec:conclusions} we give our conclusions. Throughout, we
assume a Chabrier IMF (\citealt{Chabrier03}) and assume $H_0 = 70$\kms\,Mpc$^{-1}$, $\Omega_{\rm{M}} = 0.30$ and $\Omega_{\Lambda}= 0.70$; in this cosmology, 1\,arcsec corresponds to 7.5\,kpc at
$z=0.8$ and 8.4\,kpc at $z=1.4$.

\section{Survey description, sample selection and observations}
\label{sec:survey}

\begin{figure} 
\centerline{
\psfig{figure=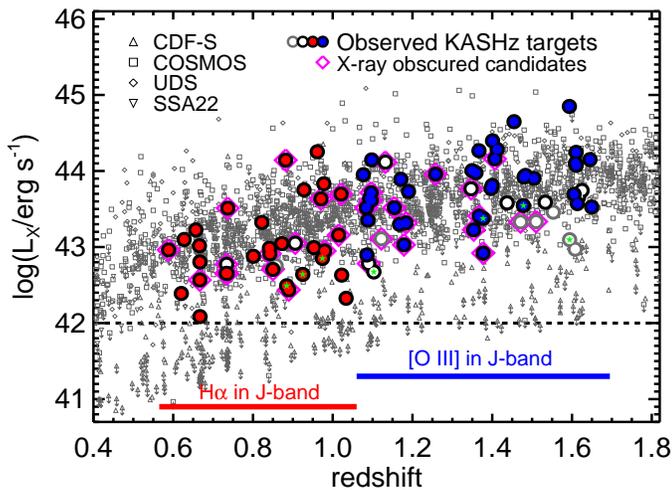,width=0.5\textwidth,angle=90}
}
\caption{Hard-band (2--10\,keV) X-ray luminosity versus redshift for
  the sources in the four deep fields covered by KASH$z$ (see
  Section~\ref{sec:sample}). The dashed line indicates our
  luminosity cut for target selection of $L_{{\rm 2-10keV}}>10^{42}$\,erg\,s$^{-1}$. The large symbols indicate the 89 X-ray AGN that
  have been observed for KASH$z$ so far. These filled and empty symbols indicate
  emission-line detections (H$\alpha$ or [O~{\sc iii}]) and no
  emission-line detections, respectively. Grey circles represent the 7 targets that are excluded from the
  analyses (i.e., resulting in a final sample of 82 targets) because a lack of an emission-line
  detection may be unphysical (Section~\ref{sec:detectionrates}). The
  stars highlight targets that have $L_{{\rm 2-10\,keV}}$ values estimated from the soft band (see
  Section~\ref{sec:sample}). Overall the targets cover nearly three orders of
  magnitude in X-ray luminosity (i.e., $L_{{\rm X}}$$\approx$10$^{42}$--10$^{45}$\,erg\,s$^{-1}$). 
} 
\label{fig:selection} 
\end{figure} 

KASH$z$ is designed to ultimately obtain spatially-resolved emission-line kinematics of $\approx$(100--200) high-redshift ($z$$\approx$0.6--3.6)
AGN. The overall aim of KASH$z$ is to provide insight into the feeding and feedback
processes occurring in the host galaxies of high-redshift AGN by
using IFS data to measure the ionised gas kinematics traced by the H$\alpha$, [O~{\sc iii}], H$\beta$, [N~{\sc ii}] and/or
[S~{\sc ii}] emission lines. The key aspect of KASH$z$ is to exploit
the unique capabilities of the multiple integral field units (IFUs) in
the KMOS instrument to perform such measurements on
larger, more uniformly selected samples of high-redshift AGN than was
possible in previous studies that used single IFU instruments. This
will make it possible to make conclusions on the overall high-redshift
AGN population and to place previous observations, of a few sources, into the
context of the parent population of AGN. Until now, this sort of
approach has only been possible for {\em low-redshift} AGN
(e.g., \citealt{Harrison14b}; \citealt{Arribas14}; \citealt{McElroy15}).

In this paper we present the first results of KASH$z$,
which focuses on the galaxy-integrated emission-line profiles of $z$$\approx$0.6--1.7 X-ray
detected AGN. In future papers we will present results on the
spatially-resolved properties of the current sample, as well as
expand the sample size when more data are obtained during our ongoing
guaranteed-time KMOS observations. 

\subsection{Parent catalogues}
\label{sec:sample}

For our target selection we make use of deep X-ray surveys performed in extragalactic
fields (COSMOS; CDF-S; UDS and SSA22). These surveys provide an efficient method for AGN selection that is largely free from host-galaxy
contamination (e.g., see review in \citealt{Brandt15}). The chosen X-ray fields are those visible
from the VLT with the deepest X-ray data available. These allow
us to uniformly select large samples of AGN that can be efficiently observed with KMOS (see
Section~\ref{sec:kmosobservations}). The four deep fields are:
\begin{enumerate}
\item Cosmic evolution survey field (COSMOS; see
  \citealt{Scoville07}). We obtain X-ray sources by combining
  the 1.8\,Ms {\it Chandra} catalogue that covers 0.9\,deg$^{2}$
  (C-COSMOS; \citealt{Elvis09}; \citealt{Puccetti09}; \citealt{Civano12}) with the wider and shallower
  1.5\,Ms XMM-{\em Newton} catalogue that covers 2.13\,deg$^{2}$
  (XMM-COSMOS; \citealt{Cappelluti09}; \citealt{Brusa10}). For sources detected in both
  surveys we use the entries in the C-COSMOS catalogue.
\item {\em Chandra} Deep Field South (CDF-S; see \citealt{Giacconi01}). We
  use the X-ray sources from the 4\,Ms {\it Chandra} catalogue that covers 464.5 arcmin$^{2}$ (\citealt{Xue11}).
\item The Subaru/XMM-Newton Deep Survey (SXDS: see
  \citealt{Furusawa08}). We use the X-ray sources from the 400\,ks
  XMM-{\em Newton} catalogue that covers 1.14\,deg$^{2}$
  (\citealt{Ueda08}). We only observe sources that are positioned inside the
  central 0.8\,deg$^{2}$ that are also covered by the near-IR Ultra Deep
  Survey (UDS; see \citealt{Lawrence07}). We use UDS as the name
  of this field hereafter.
\item The SSA22 $z=3.09$ protocluster field (see
  \citealt{Steidel98}). We use the X-ray sources from the 400\,ks {\em
    Chandra} catalogue that covers 330\,arcmin$^{2}$ (\citealt{Lehmer09}). 
\end{enumerate}

To define our sample we primarily make use of the hard-band fluxes ($F_{\rm 2-10keV}$) for each of
the X-ray sources. We also make use of the soft-band fluxes ($F_{{\rm
    0.5-2keV}}$) when there is no hard-band detection and
for calculating the $F_{\rm 2-10keV}$/$F_{{\rm 0.5-2keV}}$ flux ratios to
select X-ray obscured candidates (see
Section~\ref{sec:obscured}). We note that throughout this work, $L_{X}$ refers to the
hard-band (2--10\,keV) luminosities. To be consistent across the
four fields, we calculate fluxes making similar assumptions to the
C-COSMOS catalogue (see \citealt{Puccetti09}); i.e., we convert quoted count
rates ($CR$) to fluxes, assuming a photon index of $\Gamma=1.4$ and
a typical galactic absorption of $N_{{\rm H}}\approx2\times10^{20}$\,cm$^{-2}$. For the SSA22
and CDF-S {\em Chandra} catalogues, we use $CR$ to flux
conversion factors of 2.87$\times$10$^{-11}$\,erg\,cm$^{-2}$ and 8.67$\times$10$^{-12}$\,erg\,cm$^{-2}$ for
the hard and soft bands, respectively. We note that these conversion
factors also take into account the conversion between the quoted
2--8\,keV energy-band values to 2--10\,keV energy-band values. For the UDS XMM-{\em Newton}
catalogue, we use the conversion factors applicable for $\Gamma$=1.4 tabulated
by \cite{Ueda08} and for C-COSMOS and XMM-COSMOS we use the quoted
flux values in \cite{Puccetti09} and \cite{Brusa10}, respectively.\footnote{We note
  that XMM-COSMOS catalogue assumes a power-law index of $\Gamma=1.7$ and
  $\Gamma=2.0$ for the hard and soft bands, respectively. However,
  this results in small differences in the calculated luminosity values from assuming $\Gamma=1.4$ (i.e.,
  $\lesssim$20\,per\,cent).} The X-ray fluxes are tabulated in
Table~\ref{tab:targets}.

We obtained archival redshifts for the sources in the fields as follows: (1) for C-COSMOS we use the
compilation in \cite{Civano12}, using spectroscopic redshifts when
available and photometric redshifts otherwise; (2) for CDF-S we only target
sources with spectroscopic redshifts as compiled by \cite{Xue11}; (3) for UDS we used
the October 2010 spectroscopic redshift compilation provided by the UDS
consortium\footnote{http://www.nottingham.ac.uk/astronomy/UDS/data/data.html}
(\citealt{Smail08}; \citealt{Simpson12}; Akiyama et~al. in prep.) and (4) for SSA22 we use the spectroscopic redshifts compiled by
\cite{Lehmer09}.\footnote{For the target SSA22-39 we used the
  spectroscopic redshift from \citealt{Saez15}.} For our observed
targets the redshifts used throughout this work, $z$, are those derived from our
measured emission-lines (i.e., $z_{\rm L}$; Section~\ref{sec:linefitting}), in preference to the archival
redshifts (i.e., $z_{\rm A}$) described above. The archival redshifts
and our emission-line redshifts for the targets presented here are tabulated in Table~\ref{tab:targets}. 

The sources in our parent catalogues are plotted in the
$L_{{\rm X}}$--redshift plane in
Figure~\ref{fig:selection}. We calculate X-ray luminosities following,
\begin{equation}
\label{eq:xraylum}
L = 4 \pi D_{L}^2 F (1+z)^{\Gamma-2},
\end{equation}
where $D_{L}$ is the luminosity distance, $F$ is
the X-ray flux in the relevant energy band, $z$ is the redshift (see above) and $\Gamma$ is the photon
index. As above, we assume $\Gamma=1.4$ for all sources. Some of the X-ray sources are detected in the soft band (0.5--2.0\,keV) but not in the hard
band (2--10\,keV) and for these sources we make use the catalogued
flux upper limits to plot them in Figure~\ref{fig:selection}. However, for the targets that we observed in this work, that do not have
a hard-band detection, i.e., seven targets that are all from the COSMOS
field (see Table~\ref{tab:targets}), we estimated hard-band fluxes by
extrapolating from the soft band assuming a power-law with
$\Gamma=1.4$ (see star symbols in Figure~\ref{fig:selection}). In all
of these cases the extrapolated hard-band fluxes were consistent with the
measured upper limits and we verified that the X-ray emission was
AGN-dominated (see below).

\subsection{Sample selection}
\label{sec:selection}

Our KASH$z$ targets were selected using the archival redshifts
and X-ray luminosities described above. For this paper, we observed
targets in the NIR $J$-band, which covers a wavelength range of
$\lambda$$\approx$1.03--1.34\,\mum. Therefore, we selected targets with
redshifts in these two ranges ranges: (1) $z$$=$1.07--1.67, for which we could observe the [O~{\sc iii}]4959,5007 emission-line
doublet or (2) $z$$=$0.57--1.05, for which we could observe the
H$\alpha$ and [N~{\sc ii}]6548,6583 emission lines. 

To avoid selecting non-AGN X-ray sources (i.e., extreme starburst galaxies), we select
sources with a measured hard-band luminosity of $L_{{\rm
    2-10\,keV}}>10^{42}$\,erg\,s$^{-1}$. We note that this X-ray
luminosity has little impact on the final selection because the X-ray
catalogues are not complete down to this luminosity for our
redshift ranges of interest (see Figure~\ref{fig:selection}). Therefore, the majority of the
observed targets (i.e., all but four) have X-ray luminosities of $L_{{\rm
    2-10\,keV}}>3\times10^{42}$\,erg\,s$^{-1}$ and will have X-ray
emission that is AGN dominated; for X-ray emission of this luminosity to be produced by star-formation
processes alone, it would require extreme star-formation rates of $\gtrsim$1900\,$M_{\odot}$\,yr$^{-1}$,
based on empirical measurements of star-forming galaxies (following
Equation [4] of \citealt{Symeonidis11} and \citealt{Kennicutt98} converted to a Chabrier IMF; also see \citealt{Lehmer10}). Four of our targets, all of which are in CDF-S, have $L_{{\rm
    2-10\,keV}}=(1$--$3)\times10^{42}$\,erg\,s$^{-1}$, which
could conceivably be produced by high levels of star formation; however, we
verified that the X-ray emission is AGN-dominated in these sources by finding that their
star-formation rates, taken from \cite{Stanley15}, are too low by a factor of $>$30 to
produce the observed hard-band X-ray luminosities (following the same
procedure as above). We note that \cite{Xue11} also classify all of our CDF-S X-ray
  targets as AGN, using a more comprehensive X-ray and
  multi-wavelength identification procedure. 

The initial KMOS data of KASH$z$ that are presented here, were obtained
during ESO periods P92--P95 (details provided in Section~\ref{sec:kmosobservations}). During these observations we obtained
KMOS data for 79 X-ray AGN that met our selection
criteria. This programme was jointly observed with the KROSS
guaranteed time observing (GTO) programme (Stott et~al. 2015) and the choice of these 79 AGN targets, drawn
from our parent sample, was dictated by those that could be observed inside the KMOS pointing
positions chosen by the KROSS team, with a slight preference to
selecting $z$$\approx$1.1--1.7 AGN for which we could observe [O~{\sc
  iii}]. In summary, the observed targets are effectively
randomly selected from the luminosity and redshift plane of the parent
sample (Figure~\ref{fig:selection}; also see Section~\ref{sec:representative} for more discussion on how representative our targets are). 

We supplemented our sample of 79 KMOS targets with archival
observations taken with SINFONI (Spectrograph for INtegral Field
Observations in the Near Infrared; \citealt{Eisenhauer03}). SINFONI
contains a single-object NIR IFU, similar to the individual KMOS IFUs (see
Section~\ref{sec:sinfoni}). We queried the ESO archive\footnote{http://archive.eso.org/eso/eso{\_}archive{\_}main.html}
for $J$-band SINFONI observations at the positions of the X-ray AGN in our
parent catalogues. This yielded 10 targets that met our selection criteria, to add to the $J$-band sample
presented in this paper. 

Overall, our current KASH$z$ sample contains 89 targets. This sample consists of 54 $z$$\approx$1.1--1.7 AGN, for which we targeted the [O~{\sc
  iii}]4959,5007 emission-line doublet and 35 $z$$\approx$0.6--1.1 AGN, for which we targeted the H$\alpha$ and [N~{\sc
  ii}]6548,6583 emission lines (see Figure~\ref{fig:selection}). The targets are listed in
Table~\ref{tab:targets} and we have used a naming convention that combines their field names and corresponding X-ray ID
in the format: ``field''-``X-ray ID''. For C-COSMOS and XMM-COSMOS the
field names are shortened to COS and XCOS, respectively. 

Out of the 89 targets presented here, 83 have a spectroscopic archival redshift and 6 targets have a photometric
redshift. All of the photometric redshift targets are from the C-COSMOS field (Table~\ref{tab:targets}). We note that two of the
spectroscopic targets are classified as having ``insecure'' redshifts (e.g.,
based on low signal-to-noise ratios or single-lines; see Table~\ref{tab:targets}) by
\cite{Xue11}.\footnote{CDFS-549 is a third source with an ``insecure''
  spectroscopic redshift in \cite{Xue11} ($z_{\rm A}$$=$1.55); however, \cite{Williams14} detected H$\alpha$ at
$z$=1.553 so we re-classify it as secure.} For our emission-line
detected targets (see Section~\ref{sec:detectionrates}), our emission-line redshifts agree with
the archival redshifts, such that $|z_{L} - z_{A}|/(1+z_{L})\lesssim0.005$, except for the following
three exceptions: COS-44 ($z_{A}=1.51$; $z_{L}=0.80$) and COS-1199 ($z_{A}=0.77$; $z_{L}=0.85$), both
of which had photometric archival redshifts, and CDFS-561
($z_{A}$$=$0.80; $z_{L}$$=$0.98) which is a spectroscopic
archival redshift quoted by \cite{Popesso09}. We discuss the detection rates
of our targets, and how they relate to the archival redshifts, in Section~\ref{sec:detectionrates}. For the emission-line
detected AGN, throughout this work, we define the redshift as
$z=z_{L}$ and for the undetected targets or non-targeted AGN in the
parent sample we use $z=z_{A}$.

\subsection{The KASH$z$ sample in context}
\label{sec:representative}

\begin{figure*} 
\centerline{
\psfig{figure=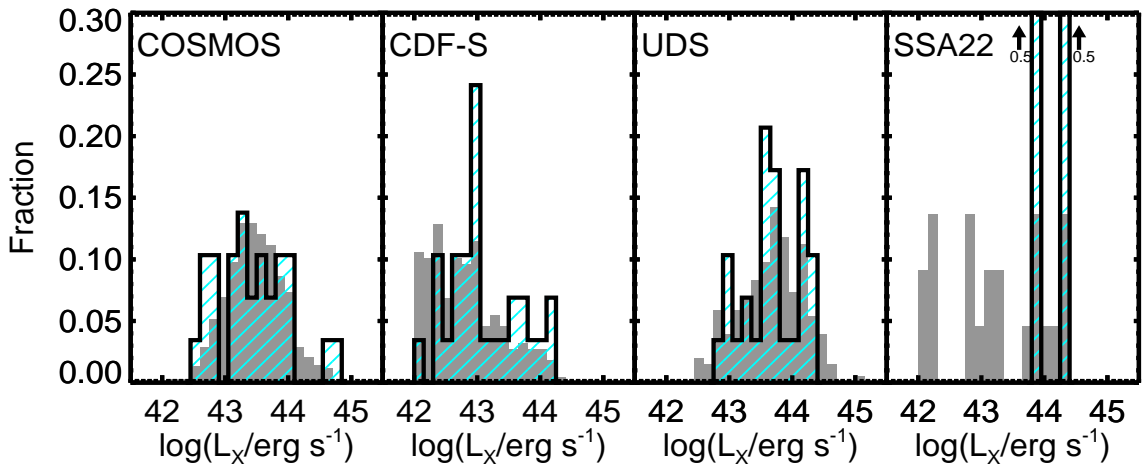,width=0.9\textwidth,angle=90}
}
\caption{The 2--10\,keV X-ray luminosity ($L_{{\rm X}}$) distributions 
  for the sources that met our selection criteria from the four deep fields described in Section~\ref{sec:sample}
  (solid histograms). Each field is shown in separate panels and we
  overlay, as hashed histograms, the targets presented in
  this work that were selected from these respective fields. With the exception of SSA22 (from which there are only two
  targets presented here), the targets have a range of X-ray luminosities that
  are representative of the parent catalogues from which they were
  selected (see Section~\ref{sec:representative}).  
} 
\label{fig:representative} 
\end{figure*} 

In Figure~\ref{fig:representative} we show histograms of the X-ray
luminosities for the X-ray sources that met our selection criteria and
of the subset of these that were observed for this work (Table~\ref{tab:targets}). This figure shows that our targets
represent the full luminosity range of the parent catalogues from
which they were selected. A two-sided Kolmogorov-Smirnov (KS)
test, yields probability values of 0.53, 0.08 and 0.96 that the two
distributions are drawn from the same distribution for COSMOS, CDF-S and UDS, respectively. Hence, there is
no evidence that the targets are {\em not} representative of the parent
population from which they were selected. For SSA22 there are only
two targets and hence it is not meaningful to perform a KS test. We
conclude that our targets are broadly representative of the X-ray AGN
population covered by the parent catalogues. It is worth noting
that, although X-rays surveys arguably provide the most uniform method
for selecting AGN, there will be some unknown fraction of luminous AGN
that are detected at other wavelengths, that are very heavily obscured
in X-rays and are not detected in these surveys
(e.g., \citealt{Vignali06}; \citealt{DelMoro15}). We defer a comparison to AGN selected
using different observational methods to future work. 

\subsection{Selecting X-ray obscured candidates}
\label{sec:obscured}

For part of this work we compare the emission-line properties of the targets
that are most likely to be X-ray obscured (i.e., $N_{H}\gtrsim10^{22}$\,cm$^{-2}$) with those that are X-ray unobscured. Due to the use of both {\em
  XMM}-Newton and {\em Chandra} catalogues in our AGN selection and due to the various depths of these catalogues, we opt to use
the simplest diagnostic possible to separate X-ray obscured and unobscured sources. We take the ratio of the hard-band
to soft-band X-ray fluxes (i.e., $F_{{\rm 2-10keV}}/F_{{\rm 0.5-2keV}}$; see
Table~\ref{tab:targets}) as a proxy for an observed power-law index,
$\Gamma_{{\rm obs}}$. For the seven targets where there
are no direct hard-band detections, we
use the hard-band flux upper limits from the original X-ray
catalogues (see Section~\ref{sec:sample}). We use a threshold of $\Gamma_{{\rm
    obs}}<1.4$, or equivalently $F_{{\rm 2-10keV}}/F_{{\rm 0.5-2keV}}>3.03$,
to select our ``obscured candidates''. Assuming a typical {\em intrinsic}
power-law index of $\Gamma=1.8$ (e.g., \citealt{Nandra94};
\citealt{George00}), this threshold corresponds to an intrinsic column density of $N_{H}\gtrsim1\times$10$^{22}$\,cm$^{-2}$, at the median redshift
of our H$\alpha$ targets (i.e., $z=0.86$), and $N_{H}\gtrsim2\times$10$^{22}$\,cm$^{-2}$, at the median redshift
of our [O~{\sc iii}] targets (i.e., $z=1.4$).  This criteria yields 32
``obscured candidates'' and 52 ``unobscured candidates'' out of the
original sample of 89. The
other 5 targets have X-ray upper limit values such that they can not be
classified.  We identify the classification of each target in
Table~\ref{tab:targets}. Reassuringly, we find that only one of the obscured candidates has an identified broad-line region
component in our data (see Section~\ref{sec:linefitting}), compared to 15 of the unobscured
candidates, as expected if X-ray obscured AGN are more likely to also have an
obscured broad line region. Furthermore, this one exception only just meets our
criteria for selecting obscured sources, i.e., it has a flux ratio of $F_{{\rm
    2-10keV}}/F_{{\rm 0.5-2keV}}=3.1$. This provides extra support that our obscured AGN criterion is reliable. 

\subsection{Selecting ``radio luminous'' AGN}
\label{sec:radiodata}

For part of this work we compare KASH$z$ targets that are
radio luminous with those that are not (Section~\ref{sec:drivers}). Therefore, we collated
the available 1.4\,GHz radio catalogues for the COSMOS, CDF-S and UDS
fields (a suitable SSA22 catalogue is not available in the literature). For COSMOS we used the 5$\sigma$ VLA catalogue of
\cite{Schinnerer10}; for CDF-S we use the 5$\sigma$ VLA catalogue presented in \cite{Miller13} and for UDS we use the 4$\sigma$
VLA catalogue in Arumugam et~al. (2015). We match the positions of our targets to the radio
positions using a 2\,arcsec matching radius. For the 87 targets in these three fields (i.e., excluding the
two SSA22 targets), we obtain 26 radio matches. We calculate radio
luminosities using the aperture-integrated flux densities and assume
a spectral index of $\alpha=-0.7$. All three catalogues have typical
sensitivities around $\approx$10\,$\mu$Jy\,beam$^{-1}$ and therefore,
we are complete to 1.4\,GHz radio luminosities of $L_{{\rm 1.4GHz}}\approx10^{24}$\,W\,Hz$^{-1}$ at the redshift ranges of interest
in this work (see Figure~\ref{fig:selection}). Therefore, for this
work we separate our targets which are ``radio luminous", i.e., with
$L_{{\rm 1.4GHz}}>10^{24}$\,W\,Hz$^{-1}$ (11 targets) from those with
$L_{{\rm 1.4GHz}}<10^{24}$\,W\,Hz$^{-1}$ (76 targets). Table~\ref{tab:targets}
highlights which targets fall into each category. Above this
luminosity threshold, sources are generally thought have AGN-dominated
radio emission and could be predominantly ``radio-loud'' sources
(e.g., \citealt{DelMoro13}; \citealt{Bonzini13}; \citealt{Rawlings15}). Furthermore, we find that 13\% of our targets are in our
``radio luminous'' category, which is broadly consistent with the observed
radio-loud fraction of $\approx$10\% for luminous AGN (e.g., \citealt{Zakamska04};
\citealt{LaFranca10}; \citealt{Kalfountzou14}). 

\subsection{Observations}
\label{sec:observations}

\subsubsection{KMOS observations and data reduction}
\label{sec:kmosobservations}
The majority of the KASH$z$ targets presented here (i.e., 79 out of the 89)
were observed using the KMOS instrument on the VLT
(\citealt{Sharples04,Sharples13}). KMOS has 24 IFUs that operate simultaneously and can be
independently positioned inside a 7.2\,arcmin diameter circular field. Each
IFU has a field of view of 2.8$^{\prime\prime}\times$2.8$^{\prime\prime}$ with a pixel scale of
0.2$^{\prime\prime}$. The 24 IFUs are fed to three spectrographs
(eight IFUs per spectrograph). For this initial KASH$z$ paper we present results of sources
observed with the $YJ$ grating that covers a wavelength range of 1.03--1.34\mum\ and has a band-centre
spectral resolution of R$\sim$3600. The KMOS observations were taken during ESO Periods 92--95,
sharing IFUs with the KROSS GTO programme,
which targeted high-redshift star-forming galaxies in the same fields
(see Stott et~al. 2015; also see Section~\ref{sec:comparison}).

Observations were carried out using ABAABAAB sequences (where A frames
are on-source and B frames are on-sky), with 600\,second integrations per position and up to
0.2\,arcsec spatial dithering between on-source frames. The targets have total on-source exposure times
of 5.4--11.4\,ks (i.e., 9--19 on-source exposures). The individual
exposure times were a result of the length of observing time with
acceptable weather conditions during the various observing runs
allocated to the GTO team and the final on-source exposure times are tabulated for individual targets
in Table~\ref{tab:targets}. The median $J$-band seeing was
0.7$^{\prime\prime}$ with 90\,per\,cent below 1.0$^{\prime\prime}$. The data were reduced using the {\sc
  esorex/spark} pipeline (\citealt{Davies13}) which flatfields,
illumination corrects, wavelength calibrates and uses observations of
standard stars, taken alongside the science frames, to flux
calibrate. Additional sky subtraction was performed using
dedicated sky IFUs. Repeated observations of targets were stacked into
the final fully reduced datacubes with a clipped
mean using the {\sc esorex} pipeline. Finally, we rebinned the cubes onto a
0.1\,arcsec pixel scale. For full details of the observations and data
reduction see Stott et~al. (2015). 

\subsubsection{SINFONI observations and data reduction}
\label{sec:sinfoni}

In addition to the primary observations using KMOS, we supplement the
KASH$z$ sample with archival observations of 10 targets, that met our
selection criteria, taken with SINFONI (see Section~\ref{sec:sample}). The observations presented here were all
observed using SINFONI's 8$\times$8\,arcsec$^{2}$
field of view, which is divided into 32 slices of width 0.25\,arcsec
with a pixel scale of 0.125\,arcsec along the slices. The observations
were carried out using the $J$-band grating which has an approximate
resolution of  $R$$\sim$3000. The observations used a variety of
observing strategies, but in all cases we were able to subtract on-sky frames from
on-target frames. We reduced the SINFONI data using the standard {\sc esorex} pipeline
(\citealt{Modigliani07}) that performs flat fielding, wavelength calibration and cube
re-construction. We flux calibrated individual data
cubes using standard star observations taken the same night as the
science observations at a similar airmass. We stacked individual cubes of the same
source by creating white-light (collapsed images), centroiding the
cubes based on these images and then performing a median stack with a
3$\sigma$ clipping threshold, rejecting cubes that could not be well centred. Three of the targets
(CDFS-51; CDFS-370 and CDFS-492) were not detected in any of the
individual cubes and for these targets, we used the offset pattern in
the headers to stack the cubes. The total on-source exposure times range
from 2.4--25.2\,ks and are tabulated for the individual targets in Table~\ref{tab:targets}.


\section{Analysis and comparison samples}
\label{sec:analyses}

In this section we describe our analyses of the galaxy-integrated
spectra for the targets presented in this work
(Section~\ref{sec:spectra}--\ref{sec:nonparams}) and we also describe our comparison samples of low-redshift AGN and high-redshift
star-forming galaxies (Section~\ref{sec:comparison}). 

\subsection{Galaxy-integrated spectra}
\label{sec:spectra}

\begin{figure} 
\centerline{
\psfig{figure=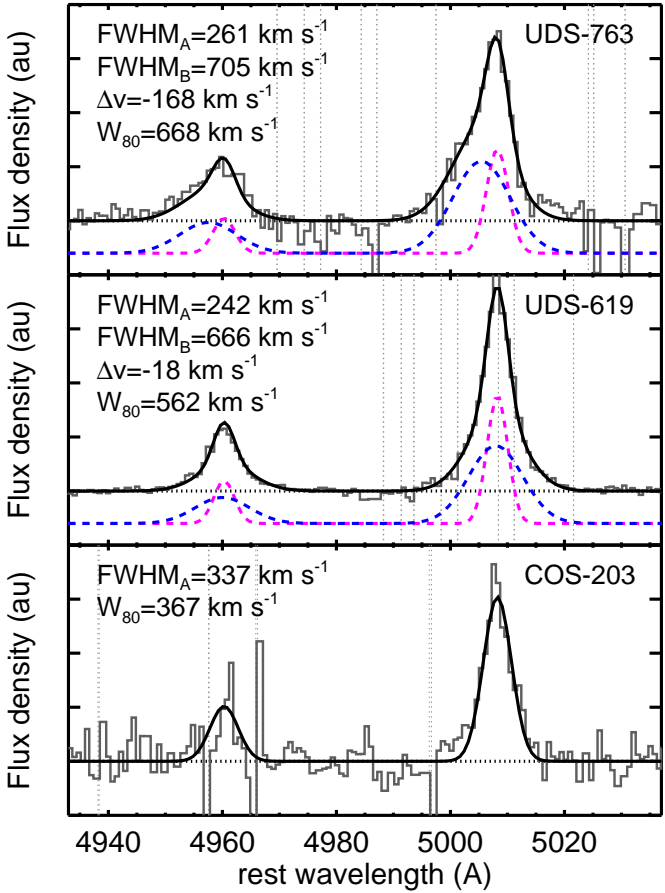,width=0.50\textwidth,angle=0}
}
\caption{Three examples of our continuum-subtracted, high signal-to-noise
  ratio, [O~{\sc iii}]4959,5007 emission-line profiles. The black curves show our fits to the emission-line profiles and the dashed curves
  show the individual Gaussian components, where applicable, with an arbitrary
  offset in the y-axis. The vertical dotted lines indicate the wavelengths of the
  brightest sky lines (\protect\citealt{Rousselot00}). These examples
  demonstrate the diversity of emission-line profiles observed in the
  sample. {\em Top:} a broad, highly asymmetric profile; {\em
    Middle:} a broad, almost symmetric profile and {\em
    Bottom:} a relatively narrow profile without a strong underlying
  broad component. Figure~\ref{fig:specoiii} presents the
  [O~{\sc iii}]4959,5007 emission-line profiles for all of the targets. 
} 
\label{fig:oiiiexamples} 
\end{figure}

\begin{figure} 
\centerline{
\psfig{figure=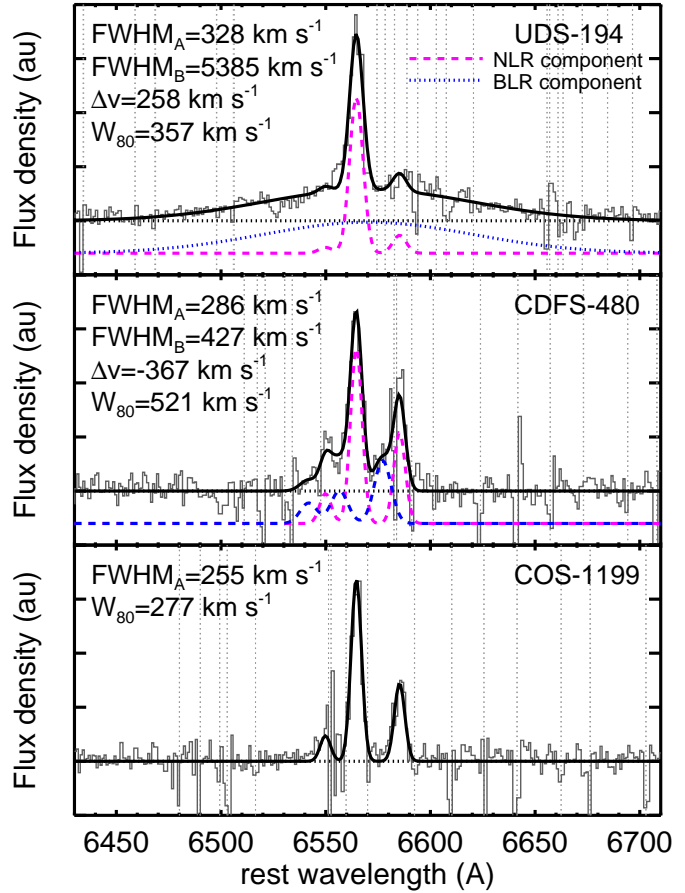,width=0.50\textwidth,angle=0}
}
\caption{Three examples of our continuum-subtracted
  H$\alpha$+[N~{\sc ii}]6548,6583 emission-line
  profiles. The black curves show our fits to the emission-line profiles and the dashed or dot-dashed curves
  show the individual Gaussian components, where applicable, with an arbitrary
  offset. The vertical dotted lines indicate the wavelengths of the
  brightest sky lines (\protect\citealt{Rousselot00}). These examples
  represent the three different types of profile fits used in this work. {\em Top:} a
  source with a BLR component in addition to the NLR component; {\em Middle:} a source with no BLR
  component but two NLR components seen in H$\alpha$ and [N~{\sc
    ii}] and {\em Bottom:} a source with a single NLR component. Figure~\ref{fig:specha} and Figure~\ref{fig:spechablr} present the
H$\alpha$+[N~{\sc ii}]6548,6583 emission-line profiles for all of the targets. 
} 
\label{fig:haexamples} 
\end{figure}

We extracted a galaxy-integrated spectrum from the KMOS and SINFONI
data cubes using the methods described below. We show example spectra in
Figure~\ref{fig:oiiiexamples} and Figure~\ref{fig:haexamples} and all
89 spectra are shown in Figure~\ref{fig:specoiii}--Figure~\ref{fig:spechablr}. 

To obtain the galaxy-integrated spectra, we initially define a galaxy
``centroid'' by creating wavelength-collapsed images from the data cubes, including both continuum and line-emission. For
three targets no emission lines or continuum were detected and we therefore assumed the centroid was at the centre of the
cube for these targets (we later exclude these three targets from our
analyses; see Section~\ref{sec:detectionrates}). For the targets CDFS-454 and CDFS-606 there are
bright continuum sources in the field of view that are spatially offset from the line-emitting
regions by $\approx$2\,arcsec and $\approx$1.2\,arcsec, respectively. For these two targets we centred on the
line emission. For each galaxy we summed the spectra
from all of the pixels within a circular aperture around the galaxy
centroids. We chose diameters to broadly match the physical
size scale of the spectra obtained with the SDSS fibres in our low-redshift
comparison sample (see Section~\ref{sec:comparison}). The median
redshift of the low-redshift SDSS comparison sample is $z$=0.14 which means
that the spectra are from a median 7.4\,kpc diameter aperture due to the
3\,arcsec fibres. For our KASH$z$ targets, the median redshift of the
[O~{\sc iii}] sample is $z$=1.4 (i.e., corresponding to 8.4\,kpc\,arcsec$^{-1}$) and the median redshift of the H$\alpha$ sample
is $z$=0.86 (i.e., corresponding to 7.7\,kpc\,arcsec$^{-1}$). Therefore, for the [O~{\sc iii}] targets we used a diameter of 0.9\,arcsec and for the
H$\alpha$ targets we used a diameter of 1.0\,arcsec. These correspond to
physical diameters of 7.6$\pm$0.8\,kpc and 7.7$\pm$0.8\,kpc,
respectively, where the uncertainties correspond to a 0.1\,arcsec
pixel scale. To assess the effect of using alternative ``galaxy wide''
  apertures, which will cover any losses of flux due to seeing, we
  also extracted spectra using 2$^{\prime\prime}$ diameter
  apertures for all sources. We discuss the corrections required to the
  emission-line luminosities in Section~\ref{sec:luminosities}. For
  the sources significantly detected in both apertures, we
  find that our emission-line width measurements (Section~\ref{sec:nonparams}) are
  consistent between both apertures within 1$\times$ the errors for 77\% of the targets and within
  2$\times$ the errors for 97\% of the targets. We defer discussion of
  how the emission-line widths change as a function of
radius to future papers.

\subsection{Emission-line fitting}
\label{sec:linefitting}

The emission-line profiles were fit with one or two Gaussian
components (with free centroids, line-widths and fluxes) and a
straight line to define the local continuum (with a free slope and normalisation). The continuum regions were defined to
be small wavelength regions each side of the emission lines being fitted. The
noise in the spectra were also calculated in these regions. The fits were performed using a
minimising-$\chi^2$ method, using the {\sc IDL} routine {\sc MPFIT}
(\citealt{Markwardt09}) and we weighted against the wavelengths of the brightest
sky lines (taken from \citealt{Rousselot00}; see
dotted lines in Figure~\ref{fig:oiiiexamples} and Figure~\ref{fig:haexamples}). Quoted line widths have been corrected for
spectral resolution, where we measured the wavelength-dependant
spectral resolution using the emissions lines in sky
spectra. We provide details of how we modelled the emission-line profiles below and tabulate the parameters
of the fits in Table~\ref{tab:targets}.

For the [O~{\sc iii}]5007,4959 emission line doublet (see examples in Figure~\ref{fig:specoiii}), we simultaneously fit the
[O~{\sc iii}]5007 and [O~{\sc iii}]4959 emission lines using
the same velocity-widths and fixing the relative centroids, using the rest-frame
wavelengths of 5008.24$\AA$\ and 4960.30$\AA$, respectively (i.e.,
their vacuum wavelengths). The flux ratio of the doublet was fixed to
be 2.99 (\citealt{Dimitrijevic07}). We initially attempted to fit this
emission-line doublet with one Gaussian component (with three free
parameters) and then with two Gaussian components (with six free
parameters) per emission line. We accepted the two-component fit if there is a significant
improvement in the $\chi^2$ values. We required
$\Delta\chi^2>15$, where this threshold was chosen to provide
the best description of the emission-line profiles across the whole
sample.\footnote{We note that absolute $\Delta\chi^2$ values are
  commonly used for model selection. For example, the Bayesian
  Information Criterion (BIC; \citealt{Schwarz78}) uses $\Delta\chi^2$ but penalises against
  models with more free parameters. This is
  defined as BIC=$\Delta\chi^2+k\ln(N)$, where N is the number of data
points and $k$ is the number of free parameters. For the fits where we
favoured two component models (including all of the H$\alpha$ fits) we
find a median of $\Delta $BIC=34 between the two and one component models, which corresponds to strong evidence in favour of the
two-component models (e.g. \citealt{Mukherjee98}).} We show the fits to the full set of [O~{\sc iii}]4959,5007
emission-line profiles in Figure~\ref{fig:specoiii}. 

For the H$\alpha$ emission-line profiles, along with the nearby [N~{\sc
  ii}]6548,6583 doublet, we follow a similar approach as for the
[O~{\sc iii}] emission-line profiles (see example fits in
Figure~\ref{fig:haexamples}). Based on the atomic transition probabilities, we fix the flux
ratio of the [N~{\sc ii}] doublet to be 3.06 (\citealt{Osterbrock06}). To reduce the degeneracy between parameters, we
couple the [N~{\sc ii}] and H$\alpha$ emission-line profiles with a
fixed wavelength separation between the three emission lines (i.e.,
by using the rest-frame vacuum wavelengths of 6549.86$\AA$,
6564.61$\AA$ and 6585.27$\AA$). We also fix the line-widths of the H$\alpha$, [N~{\sc
  ii}]6548 and [N~{\sc ii}]6583 emission-line components to be the
same. The common line-width, overall centroid and the individual fluxes of H$\alpha$ and the [N~{\sc
  ii}] doublet are free to vary. We note that the emission-line coupling is only applied for
narrow-line region (NLR) components and is not applicable for
broad-line region (BLR) components that are not observed in [N~{\sc
  ii}] emission. For these H$\alpha$ targets, we have two situations
to consider; those which exhibit a BLR component and whose which do not.
\begin{enumerate}
\item For the sources where a broad component is seen in the H$\alpha$
  emission-line profile but not in the [N~{\sc ii}] emission-line
  profile, these are the BLR or ``Type 1'' sources. We note that all
  seven BLR components that we identify have a full-width at half
  maximum (FWHM) $>$2000\,km\,s$^{-1}$ (see Table~\ref{tab:targets}), further indicating
  these are true BLRs. For these, we fit one Gaussian component to the
  [N~{\sc ii}]6548,6583 emission-line doublet and two
  Gaussian components to the H$\alpha$ emission line. The narrower of
  the H$\alpha$ components is coupled to the [N~{\sc ii}] emission as described above. The
  broader of these components; i.e., the BLR component, has a centroid, line-width and a flux that are
  free to vary. Overall, for these sources, there are seven free
  parameters in the fits and an example is given in the top panel of
  Figure~\ref{fig:haexamples}. 
\item For sources without a BLR component (i.e., the ``Type 2''
  sources) we first fit a single Gaussian component to the H$\alpha$
  emission line and [N~{\sc ii}] doublet, which are coupled as
  described above. We then add a second Gaussian
  component to all three emission lines. This is coupled in the same
  way, but with the additional constraint that we force the [N~{\sc ii}]/H$\alpha$ flux ratio
  of the broader Gaussian components to be $\gtrsim$2$\times$ that of the
  narrower component. This follows \cite{Genzel14}, who found that this was typical for high-redshift galaxies and AGN. This extra
  constraint is required to prevent considerable degeneracy in the
  fits that can lead to unphysical results. As for the [O~{\sc iii}]
  emission-line profiles, we accept the two component Gaussian fit if there is a significant
improvement in the $\chi^2$ values, i.e., $\Delta \chi^2>15$. Examples
of a one and two Gaussian component fit are
  shown in Figure~\ref{fig:haexamples}.
\end{enumerate}
The coupling between the [N~{\sc ii}] and H$\alpha$
emission-line profiles is a requirement of the restricted
signal-to-noise of the observations, to avoid a high-level of degeneracy, and is not necessarily
physical. The same coupling approach is often followed for high-redshift
galaxies and AGN, that are inevitably subject to limited signal-to-noise
observations (e.g., \citealt{Swinbank04}; \citealt{Genzel11,Genzel14}; Stott et~al. 2015),
and makes the underlying assumption that the emission lines are being
produced by the same gas, undergoing the same kinematics (see
discussion on this in Section~\ref{sec:outflowsha}). However, our
emission-line profile models appear to be a good description of the
data and are sufficient for our purposes of measuring the H$\alpha$
emission-line widths and luminosities and the [N~{\sc ii}]/H$\alpha$ flux ratios. We show the full set of fits to the H$\alpha$
emission-line profiles in Figure~\ref{fig:specha} and
Figure~\ref{fig:spechablr} and tabulate the parameters in
Table~\ref{tab:targets}. 

To calculate the uncertainties on the derived parameters, we follow the
same procedure for both the [O~{\sc iii}] and H$\alpha$$+$[N~{\sc
  ii}] emission-line profiles. We use our best-fit models to
generate 1000 random spectra by adding random noise to these fits, at the level measured in
the original spectra, and then re-fit and re-derive the parameters for
these random spectra. The quoted uncertainties are from the average of the
16th and 84th percentiles in the distribution of the parameters of
these random fits. We add an additional uncertainty of 30\% when plotting emission-line luminosities, to account for an estimated systematic
uncertainty on the absolute flux-calibration of the data cubes. 

\subsection{Measuring the overall emission-line velocity widths}
\label{sec:nonparams}

A key aspect of this work is to characterise the overall velocity
widths of the emission-line profiles. Our spectra have a range in
emission-line profile shapes and signal-to-noise values (see
Figure~\ref{fig:specoiii}--Figure~\ref{fig:spechablr}). Therefore, it
is most applicable to characterise the velocity widths with
a single non-parametric measurement that is independent of the number
of Gaussian components used in the best-fit emission-line models (see Section~\ref{sec:linefitting}). Furthermore, many
studies at low redshift use non-parametric definitions to describe the
very complex emission-line profiles (e.g., \citealt{Liu13b}; \citealt{Harrison14b}; \citealt{McElroy15}) and it is useful to be able to
compare to these studies. Therefore, we use the line-width definition,
$W_{80}$, which is the velocity width that contains 80\,per\,cent of
the emission-line flux; i.e., $W_{80}=v_{90}-v_{10}$, where $v_{10}$
and $v_{90}$ are the 10th and 90th percentiles, respectively. For a single Gaussian
  component $W_{80}=1.09\times$FWHM. These $W_{80}$ values are measured from
  the models of the emission-line profiles described in
  Section~\ref{sec:linefitting} and are tabulated in
  Table~\ref{tab:targets}. For this work, we are only interested in the kinematics of
  the host-galaxy gas and therefore, for the 7 targets with a BLR component, the $W_{80}$ measurements only refer to the
NLR emission (i.e., only the NLR components were used to calculate $W_{80}$). 

\subsection{Comparison samples}
\label{sec:comparison}

A key aspect of this work is to compare our high-redshift
AGN targets to high-redshift star-forming galaxies and low-redshift
AGN. This enables us to assess the evolution in the prevalence of
outflows observed in AGN and also to compare high-redshift galaxies that do and do not
host X-ray detected AGN. In the following sub-sections we describe how we constructed
our comparison samples. 

\subsubsection{Low-redshift AGN comparison sample}
For a low-redshift AGN comparison sample we make use of the catalogue provided by
\cite{Mullaney13}. This sample contains emission-line profile fits to
$\approx$24,000 $z<0.4$ optically-selected AGN from the SDSS
spectroscopic database. AGN are identified based on their [O~{\sc
  iii}]/H$\beta$ and H$\alpha$/[N~{\sc ii}] emission-line
ratios (following e.g., \citealt{Kewley06}) or the identification of a
BLR component. We take the 24,258 AGN in the \cite{Mullaney13}
catalogue, but reject the 37 sources where the [O~{\sc iii}]
emission-line profile fits have FWHM$=$4000\,km\,s$^{-1}$, which
signifies that the fits failed for these sources (\citealt{Mullaney13}). This leads to a final sample of 24,221 $z<0.4$
optically selected AGN. 

\cite{Mullaney13} fit the [O~{\sc iii}] emission-line
profiles of the low-redshift AGN with one or two Gaussian components, following a very similar
procedure to that adopted here (Section~\ref{sec:linefitting}). Following Section~\ref{sec:nonparams},
we measure $W_{80}$ for these sources from the [O~{\sc iii}] emission-line
profile fits, correcting for the wavelength-dependent SDSS spectral
resolution. Due to the nature of the fitting routine in
\cite{Mullaney13}, the H$\alpha$ emission-line profile fits are coupled to the [O~{\sc iii}] emission-line profile
fits. We are unable to replicate this method for our KASH$z$ H$\alpha$
targets because we do not have simultaneous [O~{\sc iii}] and H$\alpha$ constraints. Therefore, to
avoid a biased comparison of line-width measurements, we do not use the H$\alpha$ emission-line fits provided by
\cite{Mullaney13}. Instead, we use the single Gaussian component emission-line fits, as
measured by the SDSS team for these targets (\citealt{Abazajian09}), to calculate the $W_{80}$ values (see
Section~\ref{sec:nonparams}). These SDSS measurements do not separate NLR emission from BLR
emission; therefore, we are required to exclude the Type~1 AGN from
this comparison sample when comparing the H$\alpha$ emission-line widths
to the KASH$z$ targets (Section~\ref{sec:outflowsha}). However, we note that the comparison is
reasonable for our Type~2 KASH$z$ H$\alpha$ targets because all but
one of the KASH$z$ Type~2 AGN are fit with a single Gaussian
component.

To obtain X-ray luminosities for the $z<0.4$ AGN sample described
above, we match the \cite{Mullaney13} sample to data release 5 of the XMM serendipitous survey
(\citealt{Rosen15}), using a 1.5\,arcsec matching radius. This
resulted in 554 matches. We calculate X-ray luminosities using the quoted 2--12\,keV fluxes, which we convert to 2--10\,keV fluxes using a
correction factor of 0.872 and convert to hard-band X-ray luminosities using
Equation~\ref{eq:xraylum}.

For part of the analysis in this work (see Section~\ref{sec:outflowsoiii}
and Section~\ref{sec:outflowsha}) we are required to construct
low-redshift comparison samples that are luminosity-matched to our KASH$z$
samples. To do this, we randomly select sources from the sample
described above to construct the following three comparison samples: (1) a low-redshift
sample of $\approx$1000 AGN (both Type~1 and Type~2) that has an [O~{\sc iii}] luminosity distribution that is the same as our [O~{\sc
  iii}]-detected KASH$z$ targets (see Section~\ref{sec:detectionrates}); (2) a low-redshift sample of
$\approx$100 AGN (both Type~1 and Type~2) that has an X-ray luminosity distribution that is the same as our [O~{\sc
  iii}]-detected KASH$z$ targets and (3) a low-redshift sample of
$\approx$500 Type~2 AGN that has a H$\alpha$ luminosity distribution that is the same as our
H$\alpha$-detected KASH$z$ targets (see Section~\ref{sec:detectionrates}). It was not possible to construct an
low-redshift sample that was X-ray luminosity matched to our KASH$z$
H$\alpha$ targets, due to the lack of X-ray luminous Type~2 AGN in
\cite{Mullaney13}. However, we note that matching by [O~{\sc iii}]
luminosity compared to matching by X-ray luminosity
does not change the conclusions presented in Section~\ref{sec:outflowsoiii}. 

\subsubsection{High-redshift star-forming galaxy comparison sample}
To construct a high-redshift galaxy comparison sample we
make use of the KROSS survey (Stott et~al. 2015). This is a KMOS
GTO survey of $z$$\approx$0.6--1.1 star-forming galaxies, which was
observed simultaneously with KASH$z$ (see Section~\ref{sec:kmosobservations}). These targets were
selected on the basis of their K-band magnitudes and $r$--$z$
colours, to create a sample of star-forming galaxies with stellar
masses of $\approx$$10^{9-11}$\,M$_{\odot}$ (see Stott et~al. 2015). This data set provides an ideal comparison sample
of galaxies for our H$\alpha$ sample of X-ray
detected AGN, at the same redshift.

The initial phases of the KROSS survey, i.e., the data obtained during ESO
periods P92--P94 and the KMOS commissioning run, contains 514 galaxies (Stott et~al. 2015). For these
KROSS galaxies, we extract galaxy-integrated spectra from the KMOS data cubes
and perform emission-line profile fits following the same methods as
performed on the KASH$z$ sample (see Section~\ref{sec:linefitting} and
Section~\ref{sec:nonparams}). We require that H$\alpha$ emission is detected
at $\ge$3$\sigma$ and the emission-line profiles are well described using
 one or two Gaussian components plus a straight-line local continuum
 (i.e., following our methods described in Section~\ref{sec:linefitting}). Based on these criteria, we end up
 with $W_{80}$ H$\alpha$ measurements for 378 of the KROSS galaxies,
 where we have also removed any X-ray AGN. In our analyses (Section~\ref{sec:outflowsha}),
 we also make use of the stellar masses for these galaxies, as
 described in Stott et~al. (2015) and note that the average stellar mass of
 the sample we have constructed is $\log (M_{\star}/M_{\odot})$=10.3.

\subsection{Emission-line profile stacks}
\label{sec:stacks}

As part of our investigation we create {\em average} emission-line profiles by
using spectral stacking analyses on the KASH$z$
targets and comparison samples (see Section~\ref{sec:outflowsoiii} and Section~\ref{sec:outflowsha}). To create the emission-line profile stacks, we
de-redshift each continuum-subtracted spectrum to the
rest frame and then normalise each spectrum to the peak flux density
of the emission-line profile fits. We then construct the stacked
average spectra by taking a mean of the flux densities at each spectral
pixel, but removing the pixels affected by strong sky-line
residuals. An uncertainty on the average at each spectral pixel is
obtained by bootstrap resampling, with replacement, the stacks 1000
times and deriving the inner 68\,per\,cent of these stacks. Our analysis is focused on the kinematics in the host
galaxies and therefore, when stacking the H$\alpha$ emission-line profiles, we do
not include any Type~1 sources (i.e., those containing an identified BLR component). We fit the stacked spectra following the
procedures described in Section~\ref{sec:linefitting}; however, due to
the increased signal to noise we include one extra free parameter when
fitting the H$\alpha$ emission-line profiles which allows the flux
ratio of the H$\alpha$ and [N~{\sc ii}] emission lines to be free for
{\em both} Gaussian components. This means that is is possible for the
[N~{\sc ii}] and H$\alpha$ emission lines to have different line
widths (i.e., $W_{80}$ values; see discussion in Section~\ref{sec:outflowsha}). 


\section{Results and discussion}
\label{sec:results}

We present the first results from KASH$z$, which is an
ongoing programme that is utilising VLT/KMOS GTO to build up a large sample of IFS data of high-redshift
AGN (see Section~\ref{sec:survey}). We have obtained new KMOS
observations and combined them with archival SINFONI observations, to
compile a sample of 89 X-ray detected AGN observed in the
$J$-band. These AGN have redshifts of $z$=0.6--1.7 and
hard-band (2--10\,keV) X-ray luminosities in the range of
$L_{{\rm X}}=10^{42}$--10$^{45}$\,erg\,s$^{-1}$ and are representative of the parent
population from which they were selected (see
Figure~\ref{fig:selection} and Figure~\ref{fig:representative}). Of the 89 targets presented here, 54 have $z$=1.1--1.7 and were targeted to
observe the [O~{\sc iii}]4959,5007 emission-line doublet and 35 have $z$=0.6--1.1 and
were targeted to observe the H$\alpha$+[N~{\sc ii}]6548,6583 emission
lines. In this paper, we present the
galaxy-integrated emission-line profiles for all 89 targets and these
are shown, along with their emission-line profile fits, in
Figure~\ref{fig:specoiii}--Figure~\ref{fig:spechablr}. In the
following sub-sections we present: (1) the
emission-line detection rates and definition of the final sample of 82
targets used for all further analyses (Section~\ref{sec:detectionrates}); (2)
the relationship between emission-line luminosities and X-ray luminosities (Section~\ref{sec:luminosities}); and
(3) the prevalence and drivers of high-velocity ionised outflows
(Section~\ref{sec:outflowprevalence}).

\subsection{Detection rates and defining the final sample}
\label{sec:detectionrates}

We detected continuum and/or emission lines in the IFS data for 86 out
of the 89 KASH$z$ targets (i.e., 97\,per\,cent) (see
Table~\ref{tab:targets} for details). Overall, 40 targets were detected in [O~{\sc iii}], 32
were detected in H$\alpha$ and 14 were detected in continuum only. One
of the reasons for a lack of an emission-line detection appears to be due to
inaccurate photometric redshifts (photometric redshifts were only used for six COSMOS
targets; see Section~\ref{sec:selection}). Of the six targets with
photometric archival redshifts, only two resulted in emission-line detections and one of these was
detected in H$\alpha$ with $z_{{\rm L}}\approx1.5$, despite an archival redshift of
$z_{{\rm A}}\approx0.8$ (see Table~\ref{tab:targets}). To ensure that a target is undetected in line emission because of an intrinsically low
emission-line flux, as opposed to an incorrect redshift or position,
for all further analyses we only consider non-detections if: (1) they
have a secure spectroscopic archival redshift (this excludes 5 non-detections) and (2) they were
detected in continuum so that we have a reliable source position to extract the
spectrum from the datacube (this excludes 2 further
non-detections). Based on these exclusions, for the remainder of this
work, we only consider 82 of the original 89 targets, which consists of: (a) 40 emission-line detected [O~{\sc iii}]
targets and 8 [O~{\sc iii}] targets that were only detected in continuum
(i.e., a 83\,per\,cent emission-line detection rate) and (b) 32 emission-line detected H$\alpha$ targets and 2
H$\alpha$ targets that were detected only in continuum (i.e., a
94\,per\,cent emission-line detection rate). Overall this results in an emission-line detection rate of
88\,per\,cent for our final sample of 82 targets.

\begin{figure} 
\centerline{
\psfig{figure=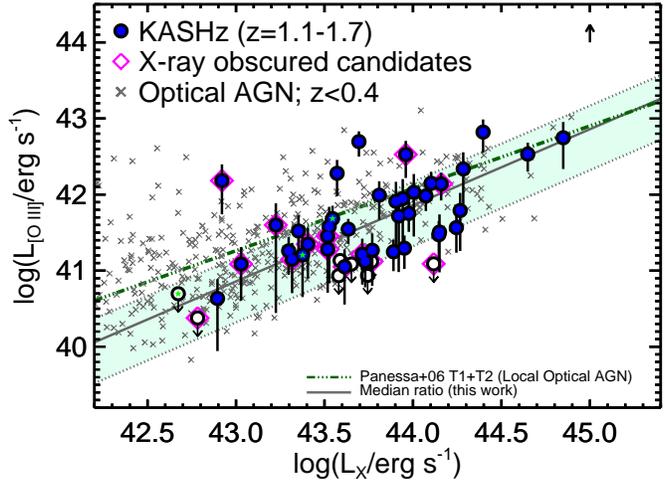,width=0.50\textwidth,angle=90}
}
\caption{Total [O~{\sc iii}] emission-line luminosity ($L_{{\rm [O~III]}}$)
  versus hard-band (2--10\,keV) X-ray luminosity ($L_{{\rm X}}$) for the
  48 $z\approx1.1$--1.7 KASH$z$ targets. Non-detections are represented as hollow
  symbols, which signify 3$\sigma$ upper limits, and stars are the same
  as in Figure~\ref{fig:selection}. The solid line shows
  the median ratio of $\log(L_{{\rm [O~III]}}/L_{{\rm X}})=-2.1_{-0.5}^{+0.3}$ for these 48 KASH$z$ [O~{\sc iii}]
  targets. The shaded region indicates the $\approx$1$\sigma$ scatter
  on this ratio (see Section~\ref{sec:luminosities}). The arrow shows the median aperture
  correction when using a $\approx$2$\times$ larger aperture (see Section~\ref{sec:luminosities}). The dot-dashed line is the
  relationship for local Seyferts and QSOs presented in
  \protect\cite{Panessa06}. We also show our $z<0.4$ optical AGN comparison sample
  (see Section~\ref{sec:comparison}). The X-ray selected KASH$z$ sample is broadly consistent
  with the low-redshift optically selected samples, but with the
  expected tendency towards lower [O~{\sc iii}] luminosities (Section~\ref{sec:luminosities}). 
} 
\label{fig:loiiilx} 
\end{figure}

We note here that the non emission-line detected sources are not
only associated with the lowest X-ray luminosity sources (see
Figure~\ref{fig:loiiilx} and Figure~\ref{fig:lhalx}). However, of the
10 targets from our final sample of 82, that were not detected in
emission lines, 5 of them are classified as X-ray obscured, 4 are classified as
unobscured and 1 is unclassified, which results in a 50--60\% obscured fraction, compared
to 33--38\% for the 72 emission-line detected targets (see
Section~\ref{sec:obscured} and Table~\ref{tab:targets}). Although we can not draw
any firm conclusions from this, it is interesting to speculate that
this provides evidence that the material obscuring the X-rays may also be responsible for the lack
of emission-line luminosity and therefore that the obscuring material
may be associated with the host galaxy (i.e., galactic-scale dust; see
e.g., \citealt{Malkan98}; \citealt{Goulding09}; \citealt{Juneau13}). 

\subsection{Emission-line luminosities compared to X-ray luminosities}
\label{sec:luminosities}

\subsubsection{[O~{\sc iii}] luminosity versus X-ray luminosity}
Both the [O~{\sc iii}] emission-line luminosity and X-ray luminosity
have been used to estimate total AGN power (i.e., bolometric luminosities) and
therefore, the relationship between these two quantities and any
possible redshift evolution, is of fundamental
importance for interpreting observations (e.g., \citealt{Mulchaey94}; \citealt{Heckman05};
\citealt{Panessa06}; \citealt{Netzer06}; \citealt{LaMassa09};
\citealt{Lamastra09}; \citealt{Trouille10}; \citealt{Lusso12}; \citealt{Berney15}). Until recently, there has been limited available
NIR spectroscopy and therefore limited measurements of the [O~{\sc
  iii}] emission-line luminosities for high-redshift, i.e.,
$z\gtrsim1$, X-ray detected AGN. In this sub-section we compare the [O~{\sc iii}] and
X-ray luminosities for our final KASH$z$ sample of 48 $z$$\approx$1.1--1.7
AGN that were targeted to observe [O~{\sc iii}] emission (see Section~\ref{sec:detectionrates}). 

In Figure~\ref{fig:loiiilx} we show [O~{\sc iii}] luminosity
versus X-ray luminosity for the KASH$z$ targets. A correlation is
observed between these two quantities, although with a large scatter,
in qualitative agreement with studies of low-redshift AGN (e.g., \citealt{Heckman05}; \citealt{Panessa06}). We note
that we observe the same trend when plotting [O~{\sc iii}] flux versus
X-ray flux and therefore the correlation is not an artifact of flux limits. For the
48 targets, we find a median luminosity ratio of $\log(L_{{\rm [O~III]}}/L_{{\rm
    X}})=-2.1^{+0.3}_{-0.5}$, where the quoted upper and lower
bounds contain the inner 66\,per\,cent of the targets, i.e.,
roughly the 1$\sigma$ scatter.\footnote{We note that we quote the range on the [O~{\sc iii}] to X-ray
  luminosity ratio for the inner 66\,per\,cent, rather than the more
  standard 68.3\,per\,cent, applicable for 1$\sigma$, because 8 out of the 48 targets are undetected (i.e., 17\%) and therefore
  we can not constrain the 15.9\,th percentile of the distribution.} To calculate the median ratio of [O~{\sc iii}]
  to X-ray luminosity we have assumed that the non emission-line detected [O~{\sc
    iii}] targets fall into the bottom 50\% of this distribution. Furthermore, to give the quoted range, we have assumed
  that they intrinsically fall in the bottom 17\,per\,cent of the $L_{{\rm
    [O~III]}}/L_{{\rm X}}$ distribution. These are not unreasonable
assumptions given that all but one of these targets have 3$\sigma$ upper limits that fall very close to, or
below, this boundary (see Figure~\ref{fig:loiiilx}). We note
  that, if we make our [O~{\sc iii}] luminosity measurements using a
  2$^{\prime\prime}$ aperture (see Section~\ref{sec:spectra}) we find a median aperture correction
  of 0.28\,dex with a scatter of 0.09\,dex. Therefore, using ``total" luminosities could increase our
  quoted $L_{{\rm [O~III]}}/L_{{\rm X}}$ ratio by $\approx$0.3\,dex
  (see Figure~\ref{fig:loiiilx}). Overall, our results
indicate that, typically, [O~{\sc iii}] luminosities are
$\approx$1\% of the X-ray luminosities, with a factor of 2--3 scatter,
for $z\approx$1.1--1.7 X-ray detected AGN.

To compare to low-redshift AGN, in Figure~\ref{fig:loiiilx}, we show the objects from our $z<0.4$
AGN comparison sample (see Section~\ref{sec:comparison}) and the
relationship found for local Seyfert galaxies and QSOs by \cite{Panessa06}. The KASH$z$
targets appear to broadly cover the same region of the $L_{{\rm [O
    III]}}-L_{{\rm X}}$ parameter space as the $z<0.4$ AGN, but
with a slight tendency towards lower [O~{\sc iii}] luminosities. This
may be partly due to the different approaches used (or not used)
  to aperture-correct the luminosities or a lack of correction for reddening to the
[O~{\sc iii}] emission-line measurements. Furthermore, the
 $z<0.4$ AGN are initially drawn from an optically selected
SDSS sample, which is [O~{\sc iii}] flux limited (see
Section~\ref{sec:comparison}), while, in contrast, the KASH$z$ is
initially X-ray selected. Indeed, a plot of [O~{\sc iii}] flux versus X-ray flux
reveals that the deficit of low $L_{{\rm [O III]}}/L_{{\rm X}}$
luminosity ratios in the $z<0.4$ sample, observed in
Figure~\ref{fig:loiiilx}, is at least partly driven by the [O~{\sc iii}] flux
limit. \cite{Heckman05} and \cite{Lamastra09} present further discussion on the
difference between optically selected and X-ray selected
samples. 

The majority of the KASH$z$ [O~{\sc iii}] targets cover a narrow X-ray
luminosity range (i.e., $L_{{\rm X}}$=2$\times$10$^{43}$--2$\times$10$^{44}$\,erg\,s$^{-1}$; see
Figure~\ref{fig:loiiilx}); therefore, we do not attempt to fit a luminosity-dependant relationship between
$L_{{\rm X}}$ and $L_{{\rm [O~III]}}$. However, we note that the local
$L_{{\rm X}}$--$L_{{\rm [O~III]}}$ relationship is found to be close to linear in log-log space, when
measured over five orders of magnitude in X-ray luminosity (i.e.,
$\log L_{{\rm [O~III]}} \propto (0.82\pm0.08)\log L_{{\rm X}}$;
\citealt{Panessa06}; also see \citealt{Lamastra09} for an even more linear
relationship for a combined sample of optical and X-ray selected AGN). At the median
X-ray luminosity of our [O~{\sc iii}] KASH$z$ sample, i.e., $\log (L_{{\rm
    X}}$/erg s$^{-1})=43.7$, the local relationship from
\cite{Panessa06} results in a luminosity ratio of $\log(L_{{\rm [O~III]}}/L_{{\rm X}})=-1.86$, which is consistent with our KASH$z$ value of
$-2.1^{+0.3}_{-0.5}$. Therefore, based on these data, we have no
reason to conclude that there is any significant difference between the $L_{{\rm
    X}}$--$L_{{\rm [O~III]}}$ relationship for our $z$$\approx$1.1--1.7 sample compared to local AGN. 

In Figure~\ref{fig:loiiilx} we observe a large scatter of a factor of
$\approx$3. This is similar to the scatter seen in local X-ray selected AGN by \cite{Heckman05}. The lack of a correction for X-ray obscuration may account for some
of this scatter; however, we note that the obscuration correction to hard-band (2-10\,keV)
luminosities, at these redshifts, will be small except in the most
extreme cases (e.g., \citealt{Alexander08b}; although see \citealt{LaMassa09}). A more
significant factor on the amount of scatter will be the effect of various
amounts of dust reddening, in the host galaxy, that will affect the
[O~{\sc iii}] emission-line luminosities (see e.g.,
\citealt{Panessa06}; \citealt{Netzer06}). Due to the lack of sufficient constraints on this
reddening effect across the sample, we do not attempt to correct for
this effect. A more interesting interpretation of the larger scatter in this
diagram is a possible lack of uniformity in the amount of gas inside the
AGN ionisation fields due to variations in opening angles and
inclinations, with respect to the host-galaxy gas. These effects will
lead to different volumes of gas being photoionised by the central
AGN. One further possible interpretation for the large
scatter observed in Figure~\ref{fig:loiiilx} is the different timescales of the accretion rates being
traced by X-ray versus [O~{\sc iii}] luminosity. Whilst the X-ray emission primarily traces nuclear
activity associated with the region very close to the accretion disk
(e.g., \citealt{Pringle81}), the [O~{\sc iii}] emission is found in NLRs, which can
extend on $\approx$0.1--10\,kpc scales (e.g., \citealt{Boroson85};
\citealt{Osterbrock06}; \citealt{Hainline13}). Therefore, a NLR that is photoionised by an AGN could
act as an isotropic tracer of the time-averaged bolometric AGN
luminosity over $\approx$10$^{4}$\,years, while in contrast, the X-ray
luminosity is an instantaneous measurement of the bolometric AGN
luminosity (see e.g., \citealt{Hickox14}; \citealt{Schawinski15}; \citealt{Berney15} for further discussion on the
relative timescales of different AGN tracers). 

\begin{figure} 
\centerline{
\psfig{figure=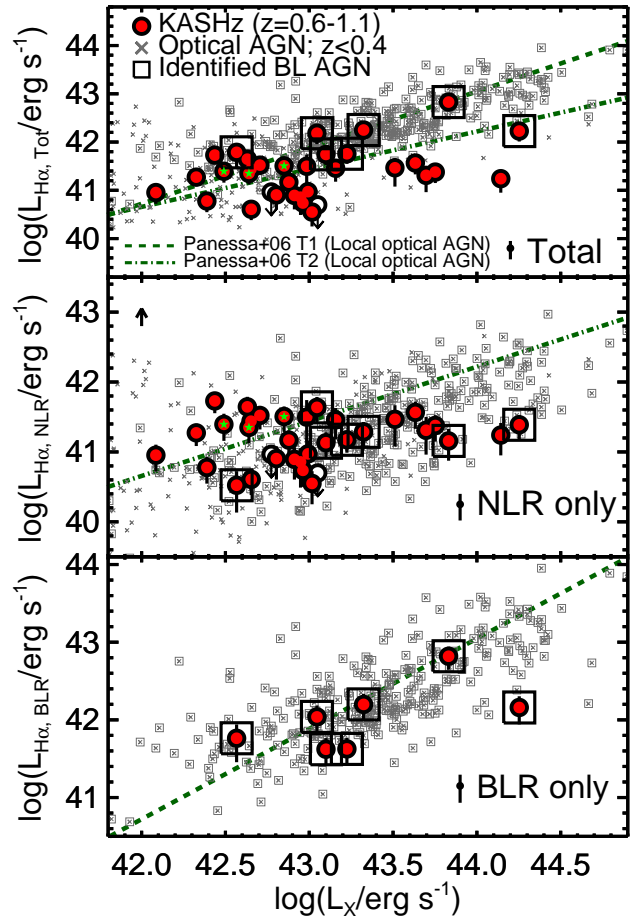,width=0.50\textwidth,angle=0}
}
\caption{H$\alpha$ emission-line luminosity ($L_{{\rm H}\alpha}$)
  versus hard-band (2--10\,keV) X-ray luminosity ($L_{{\rm X}}$) for the
  34 $z$$\approx$0.6--1.1 KASH$z$ H$\alpha$ targets. In the top panel we show the total
  emission-line luminosities, in the middle panel we show the
  luminosities where the BLR components have been subtracted (i.e., the
  NLR luminosities) and in the bottom panel we show the BLR
  luminosities, where applicable. Non-detections are represented as hollow
  symbols, which signify 3$\sigma$ upper limits, and the stars are the
  same as in Figure~\ref{fig:selection}. For clarity, the median
  uncertainty on the line luminosities is shown at the bottom of each
  panel. The arrow in the second panel shows the median aperture
  correction when using a $\approx$2$\times$ larger
  aperture (see Section~\ref{sec:luminosities}). The dashed and
  dot-dashed lines show the luminosity-dependent relationships for
  Type~1 (T1) and Type~2 (T2) local AGN presented in
  \protect\cite{Panessa06}. We also show our $z<0.4$ optical AGN
  comparison sample (see Section~\ref{sec:comparison}). } 
\label{fig:lhalx} 
\end{figure} 

\subsubsection{H$\alpha$ luminosity versus X-ray luminosity}
The H$\alpha$ emission-line luminosity is a very common tracer of
star-formation rates in local and high-redshift galaxies (e.g., see
\citealt{Kennicutt98}; \citealt{Calzetti13}). However, this is
complicated in AGN host-galaxies for two main reasons. Firstly, H$\alpha$
emission is also produced in the sub-parsec scale BLRs around SMBHs, which are most likely to be
photoionised by the central AGN and, secondly, the gas in the kpc-scale NLR of AGN
can also be photoionised by the AGN (e.g., \citealt{Osterbrock06}). The relative contribution of AGN versus
star-formation to producing the total H$\alpha$ luminosity ($L_{{\rm H\alpha}}$) will
significantly affect how well correlated $L_{{\rm H\alpha}}$ is with
tracers of bolometric AGN luminosity, such as the hard-band X-ray luminosity
($L_{{\rm X}}$).  In this sub-section we compare the H$\alpha$ and
X-ray luminosities for our final KASH$z$ sample of 34 $z$$\approx$0.6--1.1
AGN that were targeted to observe H$\alpha$ emission (see Section~\ref{sec:detectionrates}). 

In Figure~\ref{fig:lhalx} we show $L_{{\rm H\alpha}}$ versus $L_{{\rm
    X}}$ for our KASH$z$ targets. In the middle and bottom panels we
separate the H$\alpha$ luminosity into that associated with the NLRs
(i.e., $L_{{\rm H\alpha,NLR}}$) and that associated with the BLRs
(i.e., $L_{{\rm H\alpha,BLR}}$), respectively (see
Section~\ref{sec:linefitting} for details of how we separate these
components). Following from our [O~{\sc iii}] analyses in the previous section, we do
not attempt to correct for X-ray obscuration or dust
reddening (although see discussion below). In Figure~\ref{fig:lhalx} we also show
our $z<0.4$ AGN comparison sample (see Section~\ref{sec:comparison})
and the relationships found for local AGN and QSOs by
\cite{Panessa06}. We note
  that, if we make our H$\alpha$ NLR luminosity measurements using a
  2$^{\prime\prime}$ aperture (see Section~\ref{sec:spectra}) we find a median aperture correction
  of 0.25\,dex with a scatter of 0.08\,dex.

In the bottom panel of Figure~\ref{fig:lhalx} it can be seen
that there is a correlation between BLR region luminosity, $L_{{\rm H\alpha,BLR}}$, and $L_{{\rm
    X}}$ for the $z<0.4$ AGN. These sources broadly follow the local
relationship observed for the Type~1 Seyferts and QSOs by
\cite{Panessa06}, i.e., $\log L_{{\rm H\alpha,T1}} \propto (1.16\pm0.07)\log L_{{\rm X}}$. Although we are limited to seven BLR sources for the
KASH$z$ sample, our targets are found to be within the scatter of the
$z<0.4$ AGN, and therefore they qualitatively follow the same relationship
as low-redshift and local AGN (see Figure~\ref{fig:lhalx}). It is not
surprising that the BLR luminosities are tightly correlated with
the X-ray luminosities (see Figure~\ref{fig:lhalx}). The BLR is
directly illuminated by the central AGN, which is also responsible for the production of X-rays
around the accretion disk (e.g., \citealt{Pringle81};
\citealt{Osterbrock06}). Furthermore, due to the size scales of
the BLR that are typically on light-days to light-months (e.g.,
\citealt{Peterson04}), the relative timescales of the accretion events
being probed by the X-ray emission and
BLR emission will be much closer compared to the
relative timescales between the X-ray emission and the kpc-scale NLR
emission (see discussion above for the [O~{\sc iii}] NLR emission). 

In contrast to the BLR luminosities, there is little-to-no evidence for a correlation
observed between the NLR H$\alpha$ luminosity and X-ray luminosity for both the KASH$z$ sample and the $z<0.4$ comparison sample (see the middle panel of
Figure~\ref{fig:lhalx}). In qualitative agreement with this for local AGN, the relationship is less steep for Type~2 AGN compared to
Type~1 AGN and QSOs by \cite{Panessa06}; i.e.,
$\log L_{{\rm H\alpha,T2}} \propto (0.78\pm0.09)\log L_{{\rm
    X}}$. However, some correlation does still exist in local
samples, whereas we do not see evidence for this in our current high-redshift sample. In addition to the timescale arguments already discussed,
this may be due to additional contributions to the NLR H$\alpha$
luminosities, in addition to photoionisation by AGN, such as by
star-formation processes (e.g., \citealt{Cresci15}). Indeed we observe that a significant
  fraction of our sample have [N~{\sc ii}]/H$\alpha$ emission-line
  ratios that could be produced by H~{\sc ii} regions (see
  Section~\ref{sec:outflowsha}). The NLR region H$\alpha$ luminosities of the KASH$z$ targets
may follow an even shallower trend than the local relationship; however, the deviation is only observed at
the highest X-ray luminosities, where we are currently limited to a lower number
of sources (see Figure~\ref{fig:lhalx}). We also note that the KASH$z$ AGN
are likely to be systematically low in $L_{{\rm H\alpha,NLR}}$ due to the lack of obscuration
correction (also see the discussion on aperture effects above). For example, the correction to the NLR
luminosities would be $\approx$0.7\,dex, assuming the
median $A_{{\rm V,H\alpha}}=1.7$ for the star-forming galaxy comparison
sample at the same redshift (Stott et~al. 2015; Section~\ref{sec:comparison}), while the average Balmer decrement of
the $z<0.4$ AGN comparison sample implies a median correction of only
$\approx$0.3\,dex for the $z<0.4$ AGN (following \citealt{Calzetti00}). Additionally, our X-ray selection compared to the optically selected comparison samples may
also provide a systematic effect towards lower line luminosities for
the high-redshift sources (see discussion above for the [O~{\sc iii}] targets).

\begin{figure} 
\centerline{
\psfig{figure=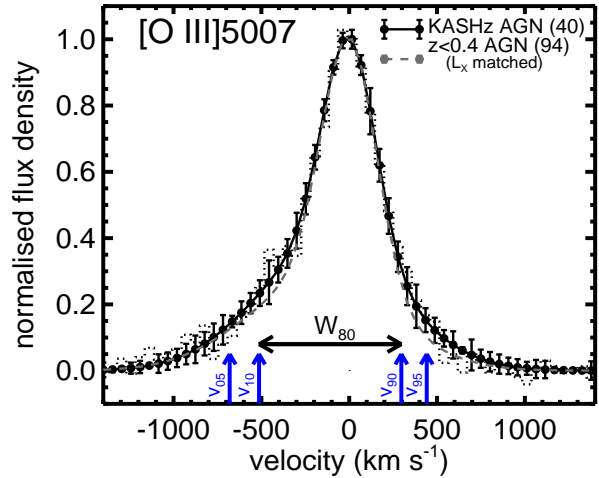,width=0.45\textwidth,angle=0}
}
\caption{Stacked [O~{\sc iii}]5007 emission-line
  profiles for the 40 [O~{\sc iii}] detected KASH$z$ targets and the
X-ray luminosity matched $z<0.4$ AGN comparison sample (see Section~\ref{sec:comparison}). The
  dotted curves show the stacked data and the dashed and solid curves are
  fits to these stacks. The upward arrows show, from left-to-right,
  the 5th, 10th, 90th and 95th percentile velocities of the KASH$z$
  stack. The overall emission-line width of $W_{80}$=810\,km\,s$^{-1}$ is also illustrated
  (see Section~\ref{sec:nonparams}). On average, the KASH$z$ AGN show
  a broad and asymmetric emission-line profile, with velocities
  reaching $\approx$1000\,km\,s$^{-1}$. The low-redshift AGN
  have a very similar average emission-line profile to the
  high-redshift AGN, for these luminosity-matched samples.} 
\label{fig:stacksoiii} 
\end{figure} 

\subsection{The prevalence and drivers of ionised outflows}
\label{sec:outflowprevalence}

A key aspect of KASH$z$ is to constrain the prevalence of ionised
outflow features observed in the emission-line profiles of
high-redshift AGN. Additionally, KASH$z$ is designed to assess which AGN and host-galaxy properties are associated with the highest
prevalence of high-velocity outflows. One of the most common
approaches to search for ionised outflows is to look
for very broad and/or asymmetric emission-line profiles in the ionised gas
species such as [O~{\sc iii}] and non-BLR H$\alpha$ components (e.g.,
\citealt{Heckman81}; \citealt{Veilleux91b}; \citealt{Mullaney13};
\citealt{Zakamska14}; \citealt{Genzel14}). For example, asymmetric emission-line profiles (most commonly a blue
wing) that reach high velocities (i.e., $\approx$1000\,km\,s$^{-1}$) are
very difficult to explain other than through outflowing material
(e.g.,\citealt{Veilleux91b}; \citealt{Zakamska14}). Furthermore, extremely broad emission-line
profiles (i.e, $W_{80}\gtrsim600$\,km\,s$^{-1}$) are very unlikely to be the result of
galaxy kinematics, but instead trace outflows or high levels of
  turbulence (e.g., \citealt{VegaBeltran01}; \citealt{Collet15}; also see discussion
in Section~\ref{sec:outflowsha}) and studies of large
samples of low-redshift AGN have shown that the gas that is producing such broad emission-line
profiles is not in dynamical equilibrium with their host galaxies
(see \citealt{Liu13b} and \citealt{Zakamska14}). 

In the following sub-sections we assess the
prevalence and drivers of ionised outflow features in the galaxy-integrated emission-line
profiles of our KASH$z$ AGN sample, following similar methods to
\cite{Mullaney13} who study $z<0.4$ AGN (see
Section~\ref{sec:comparison}). More specifically, we investigate the
distributions of emission-line velocity widths of individual sources, in combination with
emission-line profile stacks. For
clarity and ease of comparison to previous studies, we separate the discussion of the $z$$\approx$1.1--1.7 [O~{\sc iii}]
sample from the $z$$\approx$0.6--1.1 H$\alpha$ sample
(these are defined in Section~\ref{sec:detectionrates}). Furthermore, the [O~{\sc iii}] emitting gas is more
likely to be dominated by AGN illumination, while the H$\alpha$
emission may also have a significant contribution from star formation
(see discussion in Section~\ref{sec:outflowsha}). We defer a detailed comparison of these
two ionised gas tracers to future papers, which will be based on
spatially-resolved spectroscopy using both emission lines for the same
targets; however, see the works of \cite{CanoDiaz12} and \cite{Cresci15} for
IFS data covering both H$\alpha$ and [O~{\sc iii}] measurements for two high-redshift AGN. 

\subsubsection{The distribution of [O~{\sc iii}] emission-line
  velocity widths}
\label{sec:outflowsoiii}

In Figure~\ref{fig:specoiii} we show the [O~{\sc iii}] emission-line
profiles, and our best-fitting solutions, for all the $z$$\approx$1.1--1.7 KASH$z$ targets. The parameters of all of the fits are provided in
Table~\ref{tab:targets}. We identify secondary broad components (following the methods described in Section~\ref{sec:linefitting}), with
FWHM$\approx$400--1400\,km\,s$^{-1}$, in the emission-line profiles
for 14 out of the 40 [O~{\sc iii}] detected targets (i.e., 35\,per\,cent). The velocity offsets of these broad components, with respect to the narrow
components, reach up to $|\Delta v|\approx500$\,km\,s$^{-1}$. We note that \cite{Brusa15} find that four out of
their eight $z\approx$1.5 X-ray luminous AGN identify a secondary
broad emission-line component at high-significance in their
[O~{\sc iii}] spectra, which is broadly consistent with our fraction given the low numbers
involved. While the fraction of broad emission-line components in our KASH$z$
sample already indicates a high prevalence of high-velocity ionised gas in high-redshift X-ray
AGN, these measurements do not provide a complete picture across all of the
targets. This is because it is very difficult to detect multiple
Gaussian components when the signal-to-noise ratio is modest; i.e., the detection of a second Gaussian component is limited to the highest
signal-to-noise ratio spectra.  Therefore, in the following
discussion, we follow two methods to overcome these challenges. Firstly we assess the average emission-line profiles using stacking
analysis (see Section~\ref{sec:stacks}) and, secondly, we use a non-parametric definition to characterise the {\em overall} line width,
(i.e., $W_{80}$ which is the width that encloses 80\,per\,cent of the
emission-line flux; see Section~\ref{sec:nonparams}).

We show the stacked [O~{\sc iii}]5007 emission-line profile for our KASH$z$ targets in Figure~\ref{fig:stacksoiii}. The overall-emission line width of this
average profile is $W_{80}$=810$^{+130}_{-220}$\,km\,s$^{-1}$ (see Section~\ref{sec:nonparams}), where the upper and
lower bounds indicate the 68\,per\,cent range from bootstrap
resampling that stack. This indicates that, on average, the ionised
gas in these AGN have high-velocity kinematics, that are not
associated with the host galaxy dynamics. Furthermore, the average
emission-line profile clearly shows a luminous blueshifted broad
wing, which reveals high-velocity outflowing ionised gas out to velocities of
$\approx$1000\,km\,s$^{-1}$. A preference for blueshifted broad
wings, compared to redshifted broad wings, has been previously been observed for both
high- and low-redshift AGN samples (e.g., \citealt{Heckman81}; \citealt{Vrtilek85b};
\citealt{Veilleux91b}; \citealt{Harrison12a}; \citealt{Mullaney13};
\citealt{Zakamska14}; \citealt{Balmaverde15}). This can be explained if the far-side of any outflowing gas, that is moving
away from the line of sight, is obscured by dust in the host galaxies (e.g.,
\citealt{Heckman81}; \citealt{Vrtilek85b}). To observe a
redshifted component, the outflow would need to be highly-inclined or extended beyond the obscuring
material (e.g., \citealt{Barth08}; \citealt{Crenshaw10}; \citealt{Harrison12a};
\citealt{Rupke13}). 

\begin{figure} 
\centerline{
\psfig{figure=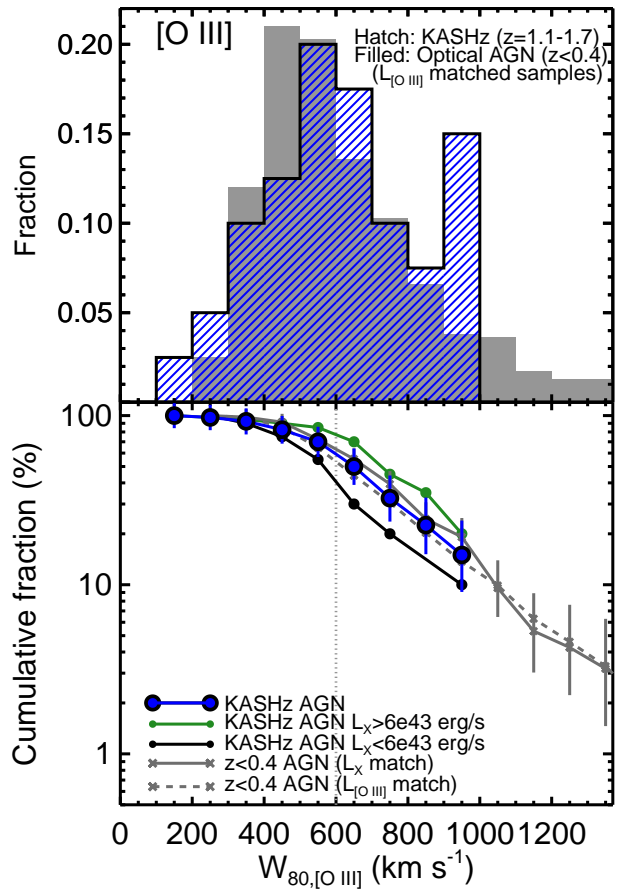,width=0.48\textwidth,angle=0}
}
\caption{{\em Top}: Histograms of the overall emission-line velocity width,
  $W_{{\rm 80,[O III]}}$, for the 40 $z\approx$1.1--1.7 [O~{\sc iii}] detected
  KASH$z$ targets (hatched) and a luminosity-matched
  sample of 1000 $z<0.4$ optical AGN (filled; see Section~\ref{sec:comparison}). The KASH$z$ AGN show a very similar
  distribution of velocities as the luminosity-matched
  low-redshift AGN sample. This is further demonstrated in the bottom panel which shows the cumulative distributions. We split the KASH$z$ sample in
  half, separating at $L_{X}=6\times10^{43}$\,erg\,s$^{-1}$, and find that high-velocity
  gas kinematics are more prevalent in higher luminosity AGN; for example,
  it is $\gtrsim$2$\times$ more likely that the higher luminosity AGN have
  $W_{80}\gtrsim$\,600\,km\,s$^{-1}$, compared to the lower-luminosity
  AGN (see vertical dotted line; Section~\ref{sec:drivers}). 
} 
\label{fig:fwhmhist} 
\end{figure}

While informative, the average emission-line profile shown in
Figure~\ref{fig:stacksoiii} hides critical
information on the underlying distribution of ionised gas kinematics
in our sample. Therefore, in Figure~\ref{fig:fwhmhist} we show the distribution of the [O~{\sc iii}] velocity-width values,
$W_{80{\rm ,[O~III]}}$, for the individually [O~{\sc iii}] detected targets. In the top panel we show the raw distribution and in
the bottom panel we show the cumulative distribution. 1$\sigma$ uncertainties
on the cumulative distribution have been calculated assuming Poisson
errors, suitable for small number statistics, following the analytical
expressions provided by \cite{Gehrels86}. We use the same uncertainty
calculations for all of the percentages presented for the remainder of
this section. We find that 50$^{+14}_{-11}$\,per\,cent
of the [O~{\sc iii}] targets have velocity-widths indicative of tracing ionised outflows or highly turbulent material, i.e.,
  $W_{80{\rm,[O~III]}}>600$\,km\,s$^{-1}$ (e.g. \citealt{Liu13b}; \citealt{Collet15}). This fraction ranges over
(42--58)\,per\,cent, when including the 8 non-detected targets (see
Section~\ref{sec:detectionrates}), for which we have no $W_{80{\rm
    ,[O~III]}}$ measurements.

To compare the prevalence of outflow features in our high-redshift AGN
sample, to low-redshift AGN, we show the distribution of
velocity-widths for our $z<0.4$ luminosity-matched comparison samples
in Figure~\ref{fig:fwhmhist} (see Section~\ref{sec:comparison}). The
luminosity-matching is performed to account for the
observed correlation between luminosity and emission-line widths (\citealt{Mullaney13}; also see
Section~\ref{sec:drivers}). We note that we obtain the same conclusions if we match
by either $L_{{\rm [O III]}}$ or $L_{{\rm X}}$. In Figure~\ref{fig:fwhmhist}, the [O~{\sc iii}] line-width distributions look indistinguishable between the KASH$z$ sample and the
luminosity-matched $z<0.4$ AGN samples. A two-sided KS test shows no
evidence that the two different redshifts have significantly different
velocity-width distributions (i.e., a $\approx$60\,per\,cent chance
that the two redshift ranges have velocity-width values drawn from
the same distribution). In Figure~\ref{fig:stacksoiii}, we further demonstrate that luminosity-matched AGN, from both redshift-ranges, have
similar [O~{\sc iii}] emission-line profiles by showing that the
stacked emission-line profiles look the same. This result implies that the prevalence of ionised outflows in low-redshift and
high-redshift AGN are very similar for AGN of the same luminosity. This is despite the fact that the average star-formation
rates of X-ray AGN are $\approx$10$\times$ higher at $z$$\approx$1
compared to $z\approx$0.2, irrespective of X-ray luminosity
(\citealt{Stanley15}). Therefore, this result provides indirect
evidence that the prevalence of these high-velocity outflows is not
greatly influenced by the level of star formation. We draw similar conclusions
for the H$\alpha$ KASH$z$ sample in Section~\ref{sec:outflowsha}.

Using our representative targets (see Figure~\ref{fig:representative} and
Section~\ref{sec:representative}), it is now possible to place some previous observations of
high-redshift X-ray AGN into the context of the overall population. For example,
the SINFONI observations of XCOS-2028 were used by
\cite{Cresci15} to present evidence of star-formation suppression by
the outflow observed in this source (see
our spectra for this source in Figure~\ref{fig:specoiii}). This
conclusion was based on their observed deficit of NLR H$\alpha$ emission, a possible tracer of star
formation (see Section~\ref{sec:luminosities} and
Section~\ref{sec:outflowsha}), in the location of the outflowing
[O~{\sc iii}] component. Based on our analyses, this source has an
emission-line velocity width of  $W_{80}\approx730$\,km\,s$^{-1}$. We find $\approx$30\,per\,cent of the
overall X-ray AGN population have these emission-line widths or
greater. Therefore, this target does not have an exceptional outflow;
however, it is yet to be determined how common the observed deficit of
NLR H$\alpha$ emission in this source is in the
overall high-redshift AGN population.

\begin{figure*} 
\centerline{
\psfig{figure=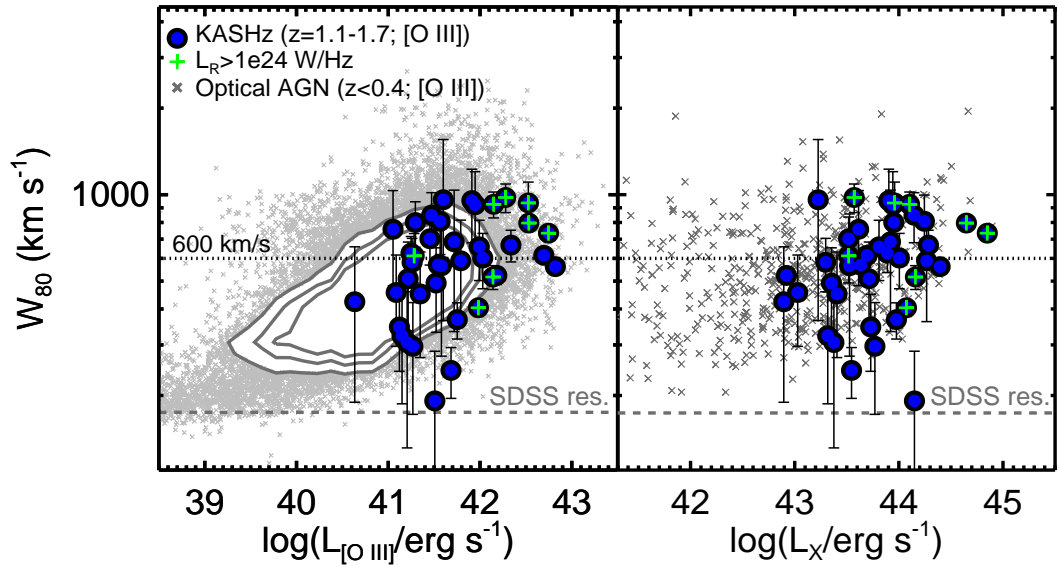,width=0.8\textwidth,angle=90}
}
\caption{Emission-line velocity width ($W_{80}$) versus [O~{\sc
    iii}] luminosity (left) and hard-band (2--10\,keV) X-ray luminosity (right) for the $z\approx$1.1--1.7 KASH$z$ sample
  (circles) and the $z<0.4$ AGN comparison sample (contours and crosses). The dashed lines show the $W_{80}$
  value for the median spectral resolution for the $z<0.4$
  sources. Although a wide distribution of velocities are observed, the most luminous AGN (either by line luminosity or X-ray
  luminosity) preferentially host extreme gas velocities. For example, all
  of the KASH$z$ AGN with $L_{{\rm[O~III]}}>10^{42.2}$\,erg\,s$^{-1}$
  have velocity-widths of $\gtrsim$600\,km\,s$^{-1}$ (see
  dotted line).} 
\label{fig:w80loiii} 
\end{figure*} 

\subsubsection{The physical drivers of the high-velocity outflows observed in [O~{\sc iii}]}
\label{sec:drivers}

It is of fundamental importance to constrain how AGN and host-galaxy properties are
related to the prevalence and properties of galaxy-wide outflows. For
example, in most cosmological models, the accretion rates of the AGN fundamentally determine the velocities and
energetics of the outflows (e.g., \citealt{Schaye15}), while some models, concentrating on individual sources, have invoked the mechanical output
from radio jets as a plausible outflow driving mechanism (e.g.,
\citealt{Wagner12}). In this sub-section we investigate the role of
AGN luminosity, radio luminosity and X-ray obscuration on the
prevalence of ionised outflows in our KASH$z$ [O~{\sc iii}] sample. 

In Figure~\ref{fig:w80loiii}, we plot the emission-line velocity width ($W_{{\rm 80,[O III]}}$) as a
function of [O~{\sc iii}] luminosity and X-ray luminosity for the
[O~{\sc iii}] detected KASH$z$ targets. Both [O~{\sc iii}] luminosity and X-ray luminosity may serve
as a tracer for the bolometric AGN luminosity, potentially on
different timescales (see Section~\ref{sec:luminosities}). We find
that the most luminous AGN (both based on [O~{\sc iii}] and X-ray) are associated with the highest
velocities, although a large spread in velocities is seen at
lower luminosities. These same trends are also seen in our
low-redshift comparison sample (see crosses in
Figure~\ref{fig:w80loiii}). To quantify the effect of AGN luminosity on the prevalence
of ionised outflows, we split the [O~{\sc iii}] detected targets into two
halves by taking the 20 ``lower'' and 20 ``higher'' X-ray luminosity
targets, resulting in a luminosity threshold of $L_{{\rm
    X}}=6\times10^{43}$\,erg\,s$^{-1}$ between the two subsets. In
Figure~\ref{fig:fwhmhist} we show the cumulative distributions of line-widths for
these two subsets. It can be seen that there is higher probability of
observing extreme ionised gas velocities in the ``higher'' luminosity sub-set. For example, 70$^{+24}_{-18}$\,per\,cent of the ``higher'' luminosity
targets have line widths of $W_{80}>$600\,km\,s$^{-1}$, while only
30$^{+18}_{-12}$\,per\,cent of the ``low'' luminosity targets reach these
line widths. A two-sided KS test indicates
only a $\approx$2\,per\,cent chance
that the two luminosity bins have velocity-width values drawn from
the same distribution. In future papers, we will be able to test this result to higher
significance as the KASH$z$ sample increases.

To first order, the results described above indicates that the highest
outflow velocities are associated with the most powerful AGN. This
result has been quoted throughout the literature for low-redshift AGN,
for both ionised outflows and for molecular outflows (e.g., \citealt{Westmoquette12}; \citealt{Veilleux13}; \citealt{Arribas14}; \citealt{Cicone14}; \citealt{Hill14}). However,
in their study of ionised ionised outflows of $\approx$24,000 AGN, \cite{Mullaney13}
found that the highest velocity outflows are more fundamentally driven
by the mechanical radio luminosity ($L_{{\rm 1.4\,GHz}}$) of the AGN, rather than the
radiative (i.e., [O~{\sc iii}]) luminosity. This result could either
be an indication that small-scale radio jets are driving high-velocity outflows, as observed in spatially-resolved studies of some low-redshift AGN (e.g.,
\citealt{Morganti05b,Morganti13}; \citealt{Tadhunter14}), or that radiatively-driven outflows are
producing shocks in the ISM which result in the production of radio
emission  (\citealt{Zakamska14}; \citealt{Nims15}; also see
\citealt{Harrison15}). For our KASH$z$ targets, we are currently limited to only eight [O~{\sc iii}]
detected targets which we can define as ``radio luminous'' (i.e.,
$L_{{\rm 1.4\,GHz}}>10^{24}$\,W\,Hz$^{-1}$; see
Section~\ref{sec:radiodata}). A higher fraction of the radio luminous
sample have high velocity line widths of $W_{80}>$600\,km\,s$^{-1}$ compared to the
non radio-luminous sources, i.e.,  6/8 or 75$^{+25}_{-30}$\,per\,cent
compared to 13/30 or 43$^{+16}_{-12}$\,per\,cent (see Figure~\ref{fig:w80loiii});\footnote{We note
  that two of our [O~{\sc iii}] detected targets do not have the
  required radio constraints to classify them as either radio luminous
  or not (see Section~\ref{sec:radiodata}).} however, the uncertainties
on these fractions are high. Furthermore, the radio
luminous sources are preferentially associated with higher X-ray
luminosity AGN and we do not currently have sufficient numbers of targets to control
for this (see Figure~\ref{fig:w80loiii}). We note that it {\em was} possible to control for AGN
luminosity, in this case [O~{\sc iii}] luminosity, for the
low-redshift study of \cite{Mullaney13}. In summary, based on the
current sample, we can not yet determine if the radio luminosity, or
the X-ray luminosity is more fundamental in driving the highest-velocity
outflows observed in high-redshift AGN.

\begin{figure} 
\centerline{
\psfig{figure=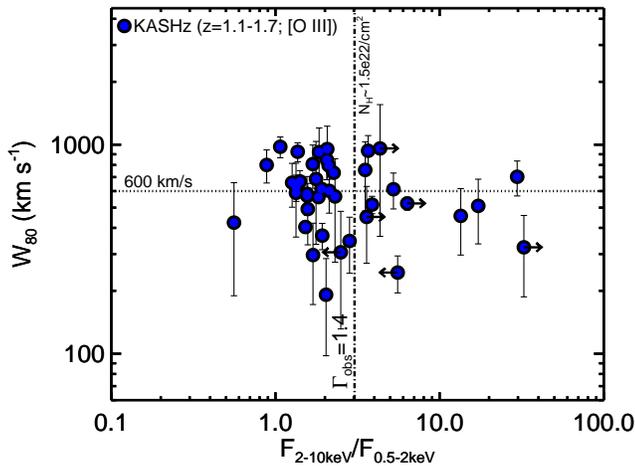,width=0.48\textwidth,angle=90}
}
\caption{Emission-line width ($W_{80}$) versus the ratio between
  2--10\,keV and 0.5--2\,keV X-ray fluxes for the 40 [O~{\sc iii}]
  detected KASH$z$ sample (circles). The X-ray axis here serves as a proxy for
  observed photon index ($\Gamma_{obs}$), and also as an estimate for
  obscuring column density (for an assumed intrinsic photon index and
  known redshift). We classify the sources
  with $F_{{\rm 2-10keV}}/F_{{\rm 0.5-2keV}}>3.03$ as ``obscured
  candidates'' (corresponding to $N_{H}\gtrsim$(1--2)$\times10^{22}$\,cm$^{-2}$;
  see Section~\ref{sec:obscured}). We find no evidence that
  high-velocity outflows are preferentially observed in the obscured X-ray AGN.
} 
\label{fig:w80gamma} 
\end{figure} 

It has been suggested that massive galaxies may go through an
evolutionary sequence where obscured AGN reside in star-forming
galaxies during periods of rapid SMBH growth and galaxy growth during
which the AGN drive outflows that drive away the enshrouding material
to eventually reveal an unobscured AGN (e.g., \citealt{Sanders88};
\citealt{Hopkins06}). Therefore, it may be expected that outflows are
more preferentially associated with X-ray obscured AGN. To test this, we compare the [O~{\sc iii}] emission-line velocities of our X-ray ``obscured
candidates'' to those ``unobscured candidates'', which we separate at
an obscuring column density of $N_{H}\approx2\times10^{22}$\,cm$^{-2}$
using a simple hard-band to soft-band flux ratio technique (see Section~\ref{sec:obscured}). In Figure~\ref{fig:w80gamma} we show
the emission-line velocity width, $W_{{\rm 80,[O III]}}$, as a function of this
flux ratio.  We find no significant difference between the two
populations; that is, we find that 45$_{-9}^{+20}$\,per\,cent of these sources have
$W_{{\rm 80,[O III]}}>600$\,km\,s$^{-1}$, compared to a similar
fraction of 54$^{+15}_{-14}$\,per\,cent for the unobscured
sources. This result implies that high-velocity ionised outflows are
certainly not uniquely, and do not appear to be preferentially,
associated with X-ray obscured AGN. This may be in
disagreement with some evolutionary scenarios for galaxy evolution; however, there may be a
population of heavily obscured AGN that are missed from even the deepest X-ray
surveys (see Section~\ref{sec:representative}) that are not present in the current sample. 

\subsubsection{The prevalence and drivers of outflows observed in H$\alpha$}
\label{sec:outflowsha}

\begin{figure} 
\centerline{
\psfig{figure=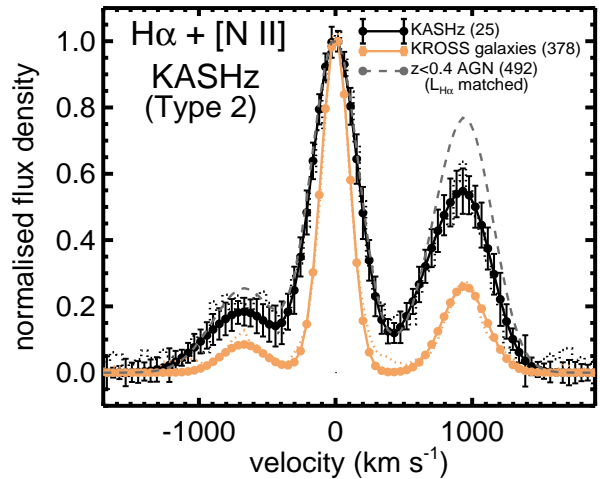,width=0.45\textwidth,angle=0}
}
\caption{Stacked H$\alpha$+[N~{\sc ii}] emission-line
  profiles for the 25 Type~2 H$\alpha$ detected KASH$z$ targets and the
  comparison samples of: (1) star-forming galaxies at the same redshift
  and (2) $z<0.4$ Type~2 AGN (see Section~\ref{sec:comparison}). The
  dotted curves show the stacked data and the dashed and solid and
  dashed curves are fits to the stacks. The KASH$z$ AGN clearly show
  much broader emission-line profiles, and higher [N~{\sc
    ii}]/H$\alpha$ emission-line ratios than the star-forming
  galaxies.} 
\label{fig:stacksha} 
\end{figure} 

In Figure~\ref{fig:specha} and Figure~\ref{fig:spechablr} we show the H$\alpha$+[N~{\sc
  ii}] emission-line profiles, and our best-fitting solutions, for all targets
in the current $z$$\approx$0.6--1.1 KASH$z$ sample. The parameters for
all of the fits are provided in
Table~\ref{tab:targets}. We detected 32 out of the 34 targets (see
Section~\ref{sec:detectionrates}) and of these 32, we identified a BLR in seven of the targets (see
Figure~\ref{fig:spechablr}). For this study, we are interested in the kinematics of the host-galaxy, or equivalently the
NLR, and not the BLR. Therefore, for these seven BLR sources the
quoted emission-line velocity widths (i.e., the $W_{{\rm 80,H}\alpha}$), are for the NLR components only (see Section~\ref{sec:linefitting}). 

In Figure~\ref{fig:stacksha} we show the stacked H$\alpha$+[N~{\sc
  ii}] emission-line profile for our 25 H$\alpha$ Type~2 AGN. The overall-emission line width of this
average profile is $W_{{\rm 80, H\alpha}}$=440$^{+50}_{-13}$\,km\,s$^{-1}$ (see Section~\ref{sec:nonparams}), where the upper and
lower bounds indicate the 68\,per\,cent range from bootstrap
resampling that stack. This velocity width is lower than the
$W_{{\rm 80, [O III]}}$=810$^{+130}_{-220}$\,km\,s$^{-1}$ observed in the stacked [O~{\sc iii}]
emission-line profile (Figure~\ref{fig:stacksoiii}). This difference
is likely to be due, in part, to the almost order of magnitude difference in the average X-ray luminosity of the
H$\alpha$ sample compared to the [O~{\sc iii}] sample (i.e.,
2$\times$10$^{43}$\,erg\,s$^{-1}$ compared to
1$\times$10$^{44}$\,erg\,s$^{-1}$; see
Figure~\ref{fig:selection}), which we have shown be a key driver for the
observed prevalence of high-velocity outflows (Section~\ref{sec:drivers}). However, as already briefly
mentioned, this may also be due to these two emission lines preferentially
tracing different regions of gas in the host galaxy, with H$\alpha$
likely to have a  larger contribution from star-forming
regions. While we currently lack systematic
IFS studies of AGN covering both [O~{\sc iii}] and H$\alpha$ emission lines, observations of some
AGN have already indicated that the NLR H$\alpha$ emission has a contribution from star-forming regions
and is not necessarily dominated by the high-velocity outflows as
is the case for [O~{\sc iii}] emission (e.g., \citealt{CanoDiaz12};
\citealt{Cresci15}). Furthermore, a strong blue wing can be observed
in the stacked KASH$z$ emission-line profile for the [N~{\sc ii}]
emission-line, which has a larger velocity
width than H$\alpha$ with $W_{{\rm 80, [N
    II]}}$=590$^{+120}_{-50}$\,km\,s$^{-1}$
(Figure~\ref{fig:stacksha}). This high-ionisation line may be more
analogous to the [O~{\sc iii}] emission, and preferentially trace
outflowing/turbulent material compared to H$\alpha$. Indeed, the difficulty in
  identifying broad-underling outflow components in H$\alpha$
  emission could be that this emission line preferentially traces
  rotation of the host galaxy and beam-smearing of this emission could dilute centrally-located broad components
  (\citealt{ForsterSchreiber14}; \citealt{Genzel14}). We are unable to
  de-couple the [N II] from the H$\alpha$ across the whole H$\alpha$ sample due
  to limited signal-to-noise ratios (see Section~\ref{sec:linefitting}) and therefore we
  may be underestimating the outflow velocities in these sources. However, H$\alpha$ emission has been
used to identify outflows in AGN host-galaxies (e.g., \citealt{Westmoquette12}; \citealt{Arribas14}; \citealt{Genzel14}) and
analysis of this emission line using these methods provides an
informative comparison to other studies.

\begin{figure}
\centerline{
\psfig{figure=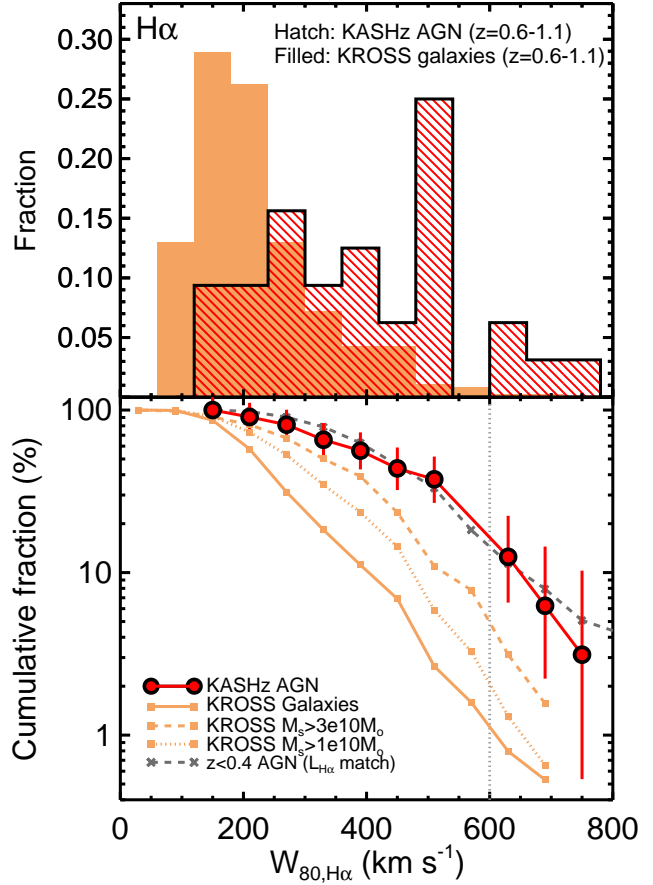,width=0.48\textwidth,angle=0}
}
\caption{{\em Top:} Histograms of the overall emission-line velocity width, $W_{{\rm 80,H\alpha}}$, for the 32
  $z$$\approx$0.6--1.1, H$\alpha$ detected KASH$z$ targets (hatched) and the $z$$\approx$0.6--1.1 comparison
  sample of star-forming galaxies from KROSS (filled; see
  Section~\ref{sec:comparison}). The AGN preferentially have higher emission-line velocities
  than the star-forming galaxies. This is further demonstrated in the
  bottom panel which show the cumulative distributions. We also show
  the KROSS sample after applying two stellar mass cuts, where the
  the $M_{\star}>3\times10^{10}$\,M$_{\odot}$ sub-sample is more comparable to the X-ray AGN host galaxies (Section~\ref{sec:outflowsha}). We also show the cumulative
  distribution of the luminosity-matched comparison sample of $z<0.4$ AGN
  (see Section~\ref{sec:comparison}) and find that the velocity distribution
  is very similar to that of our high-redshift AGN.}
\label{fig:fwhmhistha} 
\end{figure}

In Figure~\ref{fig:fwhmhistha} we show the distribution of the H$\alpha$
line-width values, $W_{80{\rm ,H\alpha}}$, for the 32 detected
targets, using the same format as above for the [O~{\sc iii}]
emission. We find that four of our KASH$z$ targets (i.e., 13$^{+9}_{-6}$\,per\,cent) have $W_{{\rm
    80,H}\alpha}>600$\,km\,s$^{-1}$, indicative of emission that is
dominated by outflowing material (see discussion at the start of this
section). We note that the range on this percentage is 12--18\,per\,cent if the two undetected targets are included, for which
we have no constraints on $W_{{\rm 80,H}\alpha}$. In agreement with the
[O~{\sc iii}] results (Section~\ref{sec:outflowsoiii}), we find no
appreciable difference between the distribution of line-widths between
the KASH$z$ sample and our low-redshift luminosity-matched comparison
sample (see Section~\ref{sec:comparison}).

\begin{figure} 
\centerline{
\psfig{figure=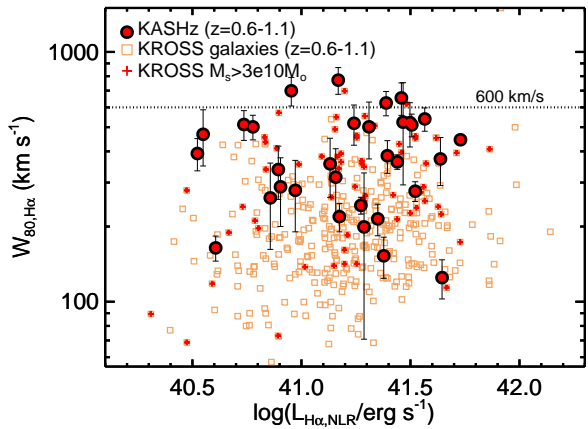,width=0.45\textwidth,angle=90}
}
\caption{Emission-line velocity width, $W_{{\rm 80,H\alpha}}$ versus narrow-line region
  H$\alpha$ luminosity ($L_{{\rm H\alpha,NLR}}$) for the KASH$z$ AGN targets (circles) and
  the KROSS star-forming galaxies at the same redshift (squares). The AGN show a similar
  distribution of $L_{{\rm H\alpha,NLR}}$, but preferentially have the
  higher velocities. $L_{{\rm H\alpha,NLR}}$ is a tracer of the
  star-formation rates for the galaxies and the AGN (although this
  will be biased upwards for the AGN; see
  Section~\ref{sec:outflowsha}). Therefore, the higher emission-line
  velocities in the AGN are not the result of higher star-formation rates.} 
\label{fig:w80halum} 
\end{figure}

In Figure~\ref{fig:fwhmhistha} we compare the H$\alpha$ emission-line
velocity distributions for our KASH$z$ targets with our comparison
sample of star-forming galaxies that are at the same redshift from KROSS (see
Section~\ref{sec:comparison}). It can clearly be seen that AGN
preferentially have higher emission-line widths than the star-forming
galaxies. For example, only 3/378 (i.e.,
0.8$_{-0.4}^{+0.8}$\,per\,cent) of the star-forming galaxy sample
reach velocity-widths of $W_{{\rm 80,H}\alpha}>600$\,km\,s$^{-1}$,
compared to 13$^{+9}_{-6}$\,per\,cent found for the AGN. This is also demonstrated in
Figure~\ref{fig:stacksha}, where we show the stacked emission-line
profiles for both samples. These results provide indirect evidence that
  the high-velocity features we observe are not pre-dominantly
  driven by star-formation. This is because X-ray AGN at these
  redshifts have average star-formation rates that are consistent with the
  global star-forming galaxy population of the same redshift (e.g.,
  \citealt{Rosario12}; \citealt{Stanley15}), and are possibly even distributed to lower
  median star-formation rates (\citealt{Mullaney15}). We further test this conclusion by plotting the
  velocity-width as a function of observed NLR H$\alpha$ luminosity
  ($L_{{\rm H\alpha,NLR}}$) for both the KASH$z$ sample and
  star-forming comparison sample in Figure~\ref{fig:w80halum}. 
  The observed NLR H$\alpha$ luminosity (i.e., excluding any BLR
  components) is a tracer of the star-formation rates in
  star-forming galaxies; however, for the
  AGN-host galaxies these will, in general be relative over-estimates due to
  the additional contribution of photoionisation by the central
  AGN. We find that the KASH$z$ AGN
  have a very similar distribution of $L_{{\rm H\alpha,NLR}}$ to our
  star-forming galaxy comparison sample, indicating that they have similar, or possibly
  lower star-formation rates. Despite this, the AGN-host galaxies preferentially have
  larger velocity widths, and a higher fraction of sources indicative
  of hosting high-velocity ionised outflows (Figure~\ref{fig:w80halum}).

In addition to investigating the role of the star formation rates, it
is also important to consider the effect of
  host galaxy masses on the H$\alpha$ emission-line widths. In the
  cases where the H$\alpha$ emission-line profiles are
  dominated by galaxy kinematics (which is the case for the
  majority of the star-forming KROSS galaxies; e.g., Stott
  et~al. 2015; Swinbank et~al. in prep), the line-widths will be driven to higher
  values in galaxies with higher stellar masses due to the increased
  velocity gradients across these galaxies. To demonstrate this,
  in Figure~\ref{fig:fwhmhistha}, we also show the cumulative
  distributions of line-widths for the star-forming galaxies when we apply increasing mass
  cuts. As expected, the higher mass galaxies tend to have the broader
  emission-line widths. We do not attempt to derive stellar masses for
  our AGN, due to the variable quality of photometric data sets
  available for our targets; however, we note that X-ray AGN appear to
typically have stellar masses of $\gtrsim3\times$10$^{10}$\,M$_{\odot}$ (e.g.,
\citealt{Mullaney12a}; \citealt{Bongiorno12}; \citealt{Aird13}; \citealt{Azadi15}). Therefore,
in Figure~\ref{fig:fwhmhistha}, we compare the star-forming galaxy
sample, but limiting it to galaxies with stellar masses
$>3\times$10$^{10}$\,M$_{\odot}$. We still find that the KASH$z$ AGN
have a higher
prevalence of the highest velocity widths, compared to this higher
mass subset of the star-forming galaxies, with only
3$^{+6}_{-1}$\,per\,cent exhibiting emission-line
velocity widths of $W_{{\rm 80,H}\alpha}>600$\,km\,s$^{-1}$.

\begin{figure} 
\centerline{
\psfig{figure=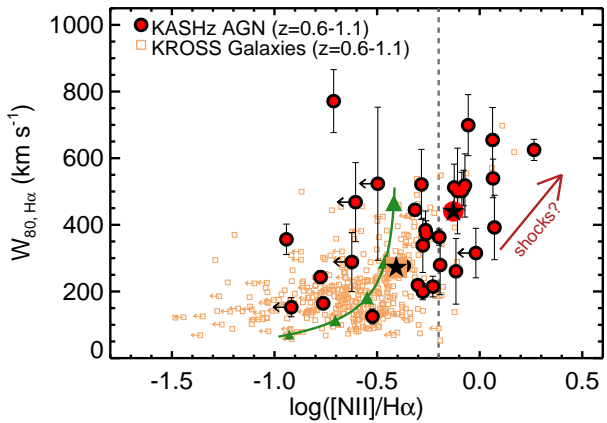,width=0.48\textwidth,angle=90}
}
\caption{Emission-line velocity width, $W_{{\rm 80,H\alpha}}$ versus
  $\log$([N~{\sc ii}]/H$\alpha_{\rm NLR}$) emission-line ratio for the KASH$z$ AGN targets (circles) and
  the KROSS star-forming galaxies (squares). The larger symbols
  containing the stars are measured from the stacked emission-line
  profiles (Figure~\ref{fig:stacksha}). The vertical dashed line indicates the
  maximum emission-line ratio expected for photoionised H~{\sc ii} regions
  (e.g., \citealt{Kewley13}). The green track shows the
  predicted trends as a function of mass (following the
  mass-metallicity and mass-velocity dispersion relations; see
  Section~\ref{sec:outflowsha}), where the triangles highlight various $\log(M_{\star})$
  values in half dex bins starting at 9.0. The positive correlation observed for most of the AGN (i.e., those with $\log$([N~{\sc ii}]/H$\alpha$)$\gtrsim$$-0.5$ may be indicative of
  outflows/shocks driving the velocity widths (see Section~\ref{sec:outflowsha}). Most
  of the star-forming galaxies roughly follow the predicted mass-driven trend
  (Section~\ref{sec:comparison}) and the stacked average is
  consistent with the mass-driven prediction at their average mass, i.e., $\log (M_{\star}/M_{\odot})$=10.3. However, they appear to follow the same trend as the AGN at the largest velocity widths ($W_{80}\gtrsim$400\,km\,s$^{-1}$).
} 
\label{fig:w80n2} 
\end{figure}

To further test the possible role of mass in driving the high emission-line
velocity widths observed in our targets, in Figure~\ref{fig:w80n2} we plot $W_{{\rm
    80,H}\alpha}$ as a function of the emission-line ratio
$\log$([N~{\sc ii}]/H$\alpha$). This emission-line ratio is a tracer of
the metallicity of star-forming galaxies (e.g., \citealt{Alloin79};
\citealt{Denicolo02}; \citealt{Kewley02}), as well as an indicator for the source of ionising
radiation (e.g. \citealt{Kewley06}; \citealt{Rich14}). Furthermore,
there is an observed relationship between mass
and metallicity and therefore an expected
relationship between this emission-line ratio and stellar mass (e.g., \citealt{Lequeux79};
\citealt{Tremonti04}; \citealt{Maiolino08}; \citealt{Stott13}). We can also make a crude prediction for the
relationship between $W_{{\rm 80,H}\alpha}$ and mass, under the
assumption that the line-width is a tracer of the stellar
velocity dispersion. Therefore, for a given stellar mass, we predict
  the position galaxies would be located in Figure~\ref{fig:w80n2} by combining: (1) the
  observed $z=0.7$ mass-metallicity relation (following \citealt{Maiolino08}) and (2) the observed mass-velocity
  dispersion relationship for massive galaxies (following
  \citealt{Bezanson15}). The majority of the star-forming galaxies
  appear to broadly follow the expected trend. Furthermore, the measurements from
  the stacked average emission-line profile (Figure~\ref{fig:stacksha}) are in agreement
  with the rough mass-driven prediction for the average mass of these
  galaxies (i.e., $\log (M_{\star}/M_{\odot})$=10.3; see Figure~\ref{fig:w80n2}). In
  contrast, the AGN typically have higher $\log$([N~{\sc ii}]/H$\alpha$)
  emission-line ratios and higher velocity widths that the
  star-forming galaxy sample (also visible in the stacked profiles; Figure~\ref{fig:stacksha}) and a positive correlation is observed between
  these two quantities. Such a positive correlation has been shown
  to be a tracer of shocks and outflows in the ISM through
  IFS observations of AGN and star-forming galaxies (e.g., \citealt{Ho14};
  \citealt{McElroy15}). Interestingly, the small number of star-forming galaxies with high line-widths (i.e., $W_{80}\gtrsim400$\,km\,s$^{-1}$) appear to
  follow the same relationship as the KASH$z$ AGN, which may
  indicate that these galaxies also have a contribution from shocks
  and/or host AGN that were not detected in the X-ray surveys. We clarify that some of the AGN targets have $\log$([N~{\sc ii}]/H$\alpha$) emission-line
    ratios that could also be photoionised by H~{\sc ii} regions (see
    Figure~\ref{fig:w80n2}) and we will explore these ideas further when exploring the
  spatially-resolved outflow kinematics and emission-line flux ratios of
  the KASH$z$ AGN in future papers.

We conclude that the highest H$\alpha$ line-widths (i.e., those
with $W_{80}>600$\,km\,s$^{-1}$ and possibly those with
$W_{80}\gtrsim400$\,km\,s$^{-1}$) are at least partially driven by ionised
outflow kinematics and/or shocks in the ISM. Furthermore, there is a increased likelihood to find these
highest ionised gas velocities in X-ray identified AGN compared to star-forming
galaxies at the same redshift that have similar star-formation
rates and masses. This is in agreement with local IFS studies that
have found  higher-velocity ionised outflows, traced by H$\alpha$
emission, in star-forming galaxies that host AGN compared to those which do not (e.g., \citealt{Westmoquette12}; \citealt{Arribas14}).


\section{Conclusions}
\label{sec:conclusions}

We have presented the first results of the ongoing KMOS AGN Survey at
High redshift (KASH$z$). The first 89 targets, which are presented
here, are high-redshift ($z$$\approx$0.6--1.7) X-ray detected AGN, with hard-band (2--10\,keV) luminosities in
the range $L_{{\rm X}}=10^{42}$--10$^{45}$\,erg\,s$^{-1}$. The
targets have a distribution of X-ray luminosities that are representative of
the parent X-ray AGN population. The majority of the targets (79) were
observed with KMOS, supplemented with archival SINFONI observations of
10 targets. All of the observations were carried out in the $J$-band
and we detected 86 (i.e., 97\,per\,cent) of the targets in continuum and/or
emission lines. However, for the analyses in this work, we excluded
seven un-reliable non-detections (e.g., due to inaccurate photometric
redshifts; Section~\ref{sec:detectionrates}). This leaves a final sample of 82
targets, for which we have presented results and discussion based on their galaxy-integrated
emission-line profiles. 

We detected 72 of the final sample (i.e., 88\,per\,cent) in
emission lines, of which 40 out of 48 targets were detected in [O~{\sc
  iii}] (the $z$$\approx$1.1--1.7 targets) and 32 out of 34 targets were detected
in H$\alpha$ (the $z$$\approx$0.6--1.1 targets). We
have explored the emission-line luminosities as a function of X-ray
luminosity for our targets and by characterising the individual emission-line profiles
and using stacking analyses, we have investigated the
prevalence and drivers of the broad and asymmetric emission-line profiles that are indicative of ionised outflows. Our main conclusions
are listed below.

\begin{itemize}
\item We find a median X-ray luminosity to [O~{\sc iii}] luminosity
  ratio of $\log(L_{{\rm [O~III}]}/L_{X})=-2.1_{-0.5}^{+0.3}$, where
  the range is roughly the 1$\sigma$ scatter. The observed relationship between these
  two quantities for our high-redshift sample is broadly consistent with that
found for low-redshift AGN and local Seyferts and QSOs (see  Figure~\ref{fig:loiiilx}). Our
  results indicate that the [O~{\sc iii}] luminosities are typically
  $\approx$1\% of the X-ray luminosities for $z\approx$1.1--1.7 X-ray
  AGN. The large scatter of a factor of $\approx$3, may be due to several
  observational effects including dust reddening, or due to intrinsic
  physical effects, such as a higher level of variability in the
  sub-pc-scale production X-ray emission compared to
  the kpc-scale production of [O~{\sc iii}] emission (Section~\ref{sec:luminosities}).
\item Our seven Type~1 H$\alpha$ targets have broad-line
  region luminosities ($L_{{\rm H\alpha,BLR}}$) that are broadly
  consistent with the correlation observed between $L_{{\rm H\alpha,BLR}}$ and
  $L_{X}$ for low-redshift and local AGN. Although limited by
    small numbers of very luminous sources, we find no evidence for
    a correlation between the narrow-line
  region H$\alpha$ luminosity and  $L_{X}$ (see
  Figure~\ref{fig:lhalx}; Section~\ref{sec:luminosities}).
\item High-velocity emission-line features are
  common in our [O~{\sc iii}] sample, with $\approx$50\,per\,cent of
  the targets exhibiting velocities indicative of being dominated
    by outflowing ionised gas or highly turbulent material (i.e., emission-line velocity widths of $W_{{\rm 80,[O
      III]}}$$>$600\,km\,s$^{-1}$; see
  Figure~\ref{fig:stacksoiii}; Figure~\ref{fig:fwhmhist} and
  Section~\ref{sec:outflowsoiii}). On average the emission-line
  profiles have a prominent blue-shifted wing, implying outflowing material. Outflowing or highly turbulent
    material that does not dominate the individual emission-line profiles could
    be even more common.
\item The high-velocity [O~{\sc iii}] kinematics are more prevalent
  for targets with higher AGN luminosities. For example, $\approx$70\,per\,cent of
  the $L_{X}>6\times10^{43}$\,erg\,s$^{-1}$ targets have [O~{\sc iii}]
  line widths of $W_{{\rm 80,[O III]}}$$>$600\,km\,s$^{-1}$, while only $\approx$30\,per\,cent of the
  $L_{X}<6\times10^{43}$\,erg\,s$^{-1}$ targets reach these line
  widths (see Figure~\ref{fig:fwhmhist} and Figure~\ref{fig:w80loiii}). Using our current sample, we are unable to determine
  the role of radio luminosity in driving this trend (Section~\ref{sec:drivers}). 
\item Based on our current X-ray detected sample, we find no evidence
  that the highest ionised gas velocities are
  preferentially associated with X-ray obscured AGN (i.e., those with
  $N_{H}\gtrsim10^{22}$\,cm$^{-2}$), compared to X-ray unobscured AGN (see
  Figure~\ref{fig:w80gamma}), in contrast to the predictions of some evolutionary
  scenarios (Section~\ref{sec:drivers}). 
  \item We compared the emission-line widths of our H$\alpha$ detected
    X-ray
    AGN targets, with a redshift matched sample of star-forming
    galaxies. Despite a similar distribution of H$\alpha$ luminosities
    (excluding the broad-line region components), and implied similar
    star-formation rates, the AGN-host galaxies exhibit a much higher
    prevalence of high ionised gas velocities (see
    Figure~\ref{fig:fwhmhistha} and Figure~\ref{fig:stacksha}). For example, $\approx$13\,per\,cent of the AGN
  have H$\alpha$ emission-line widths of $W_{80}>$600\,km\,s$^{-1}$,
  whilst only $\approx$1\,per\,cent of the star-forming galaxy sample reach these line
  widths (Section~\ref{sec:outflowsha}).
\item For both the [O~{\sc iii}] and H$\alpha$ KASH$z$ targets, we find no significant difference between the distribution of
  velocity widths for our high-redshift AGN sample and
  luminosity-matched comparison samples of $z<0.4$ AGN. Under the
  assumption that the most extreme ionised gas velocities are
  associated with outflows, to first
  order, this implies that it is just as likely to find an ionised
  outflow of a certain velocity in low-redshift AGN as in
  high-redshift AGN of the same luminosities (see Figure~\ref{fig:fwhmhist} and
  Figure~\ref{fig:fwhmhistha}). This is despite the order-of-magnitude
  global decrease in average star-formation rates towards the lower
  redshift sample (see Section~\ref{sec:outflowsoiii} and Section~\ref{sec:outflowsha}). 
\end{itemize}

Based on our systematic study of a representative sample of
high-redshift X-ray detected AGN, we have evidence that high-velocity
ionised outflows are prevalent, in qualitative agreement with theoretical predictions
of galaxy formation models. These features appear to be most common in the most powerful
AGN, and are equally prevalent in the high-redshift Universe as in the
low-redshift Universe for AGN of the same luminosity. Our
  analyses focused on searching for emission-line profiles which are
   dominated by these extreme gas kinematics and lower level outflows
  or highly turbulent material could be even more common. Due to a higher fraction of galaxies hosting the most luminous AGN at
higher redshifts, our results imply that the most extreme ionised outflows are more prevalent in
  high-redshift galaxies. In future papers, we will present results
  based on the spatially-resolved kinematics of multiple emission
  lines which will reveal information on the sizes and
    morphologies of the high-velocity gas, enable us to disentangle
    outflows from galaxy kinematics, and to measure the energetics of
the outflows (following e.g., \citealt{Nesvadba08};
\citealt{Harrison12a, Harrison14b}; \citealt{CanoDiaz12};
\citealt{Cresci15}). Furthermore, as our sample size grows, we will
be able to make more definitive tests on the drivers and impact of ionised outflows in large
samples of representative AGN.


\subsection*{Acknowledgements}
We thank the referee for their constructive comments. We acknowledge the Science and Technology Facilities Council
(CMH, DMA, JPS, AMS, RGB and RMS through grant code ST/L00075X/1) and the Leverhulme Trust
(DMA). JRM acknowledges support from the University of Sheffield via
its Vice-Chancellor Fellowship scheme. JPS acknowledges support from a
Hintze Research Fellowship. FEB acknowledges
support from CONICYT-Chile (Basal-CATA PFB-06/2007, FONDECYT 1141218,
``EMBIGGEN'' Anillo ACT1101), and Project IC120009 ``Millennium
Institute of Astrophysics (MAS)'' funded by the Iniciativa
Cient\'{i}fica Milenio del Ministerio de Econom\'{i}a, Fomento y
Turismo. We thank the members of the
KROSS team for access to their data and for
assisting with KASH$z$ observations. We also thank Holly Elbert
and Timothy Green for carrying out some observations. We thank F. Stanley
for providing star formation rate measurements and Ian Smail for
useful discussions.  


\vspace{-1cm}
\bibliographystyle{mn2e}

 
\appendix
\section{Emission-line profiles and tabulated data}

In this Appendix we provide the galaxy-integrated spectra,
emission-line profile fits and tabulated data for all 89 of the
KASH$z$ targets presented in this work.

\begin{figure*} 
\centerline{
\psfig{figure=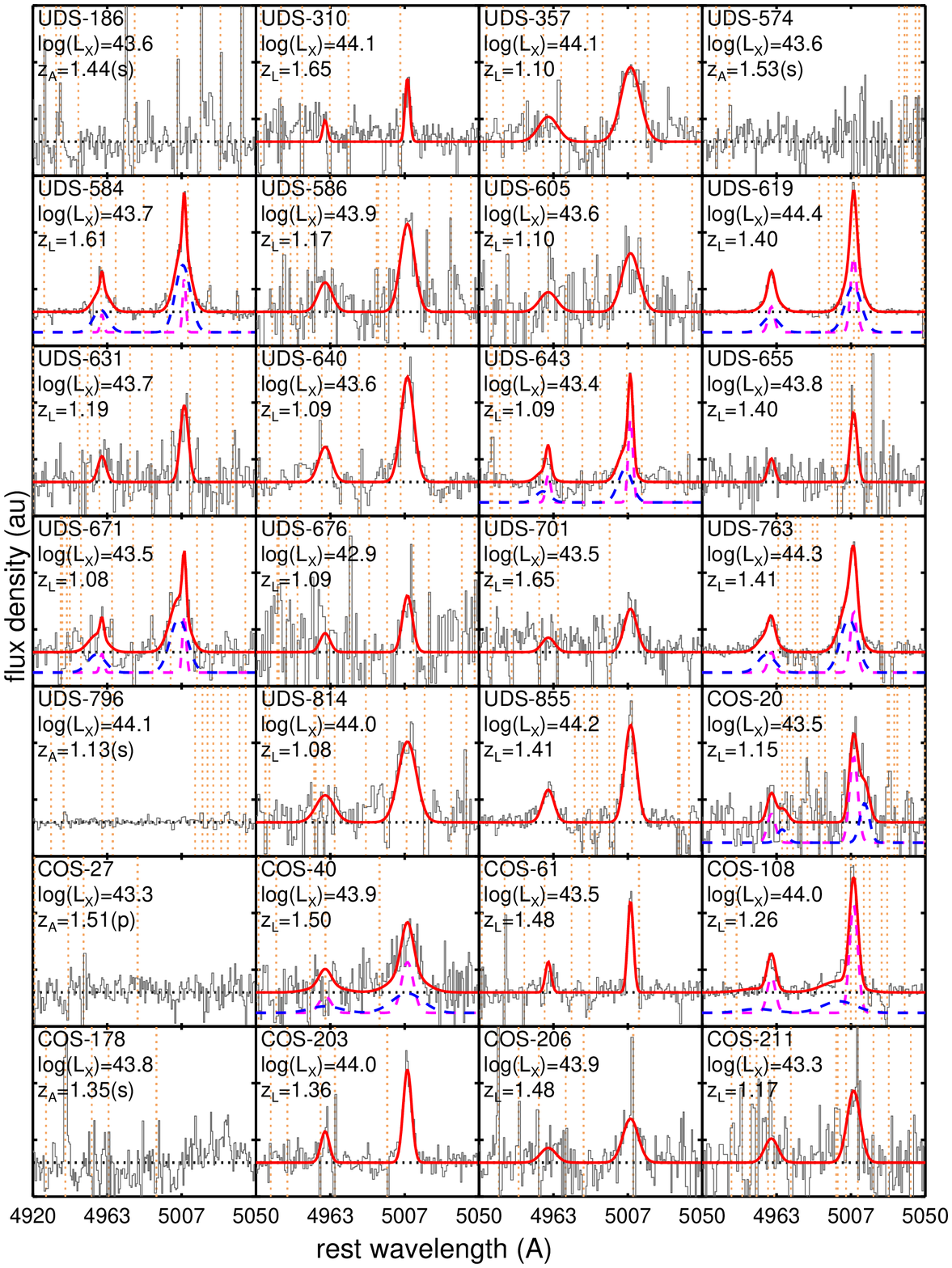,angle=0,width=0.9\textwidth}
}
\caption{Galaxy-integrated spectra, shifted to the
  rest frame, around the [O~{\sc iii}]4959,5007 emission-line
  doublet, in arbitary flux density units, for the $z\approx$\,1.1--1.7
  targets. For descriptions of the different curves see
  Figure~\ref{fig:oiiiexamples}. The vertical dotted lines indicate the wavelengths of the
  brightest sky lines (\protect\citealt{Rousselot00}). In each panel we
  also identify the target name, the logarithm of the hard-band X-ray luminosities
  (erg\,s$^{-1}$) and the redshifts (Table~\ref{tab:targets}).} 
\label{fig:specoiii} 
\end{figure*} 

\setcounter{figure}{0}

\begin{figure*} 
\centerline{
\psfig{figure=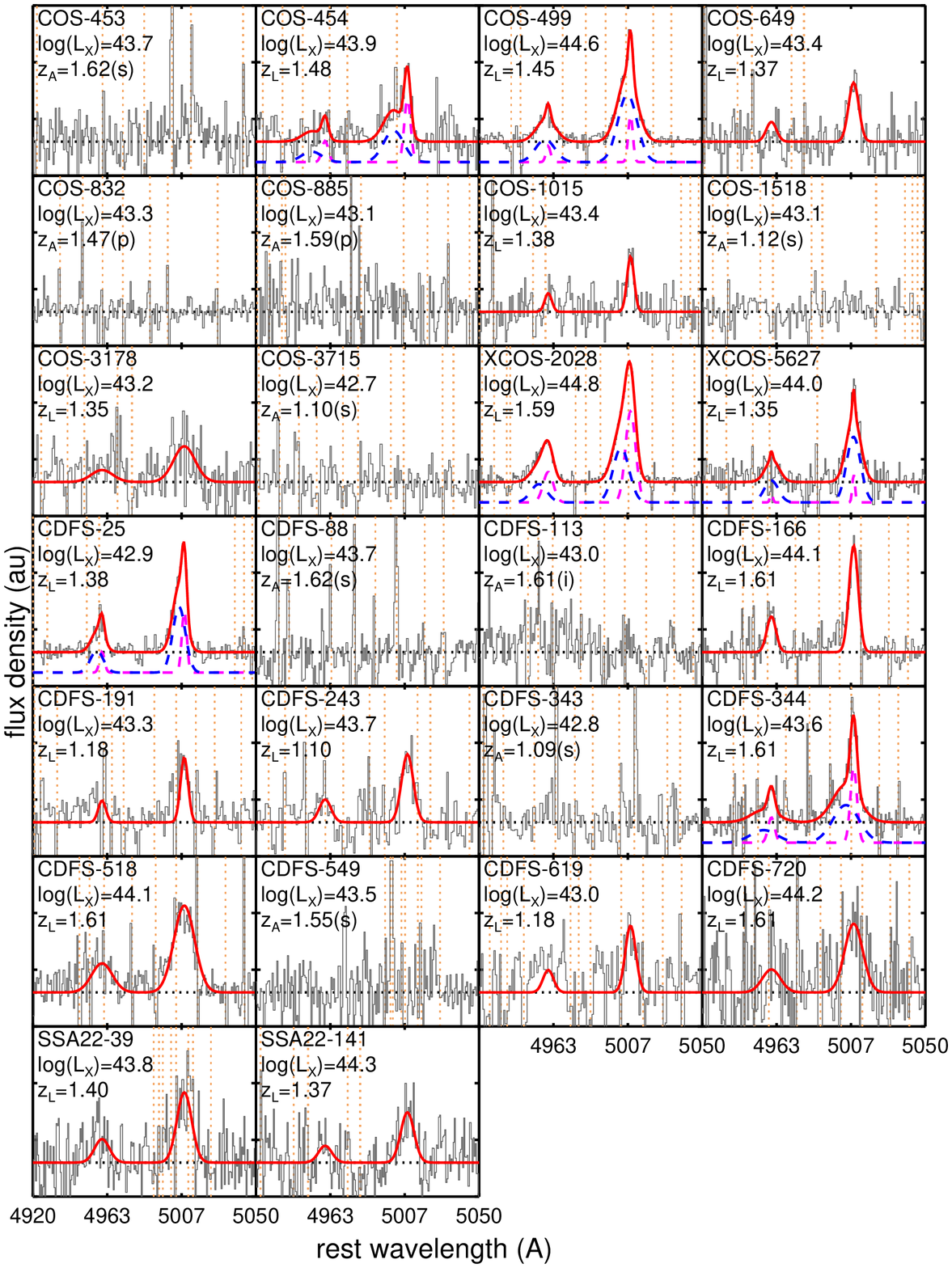,angle=0,width=0.9\textwidth}
}
\caption{continued} 
\end{figure*} 

\begin{figure*} 
\centerline{
\psfig{figure=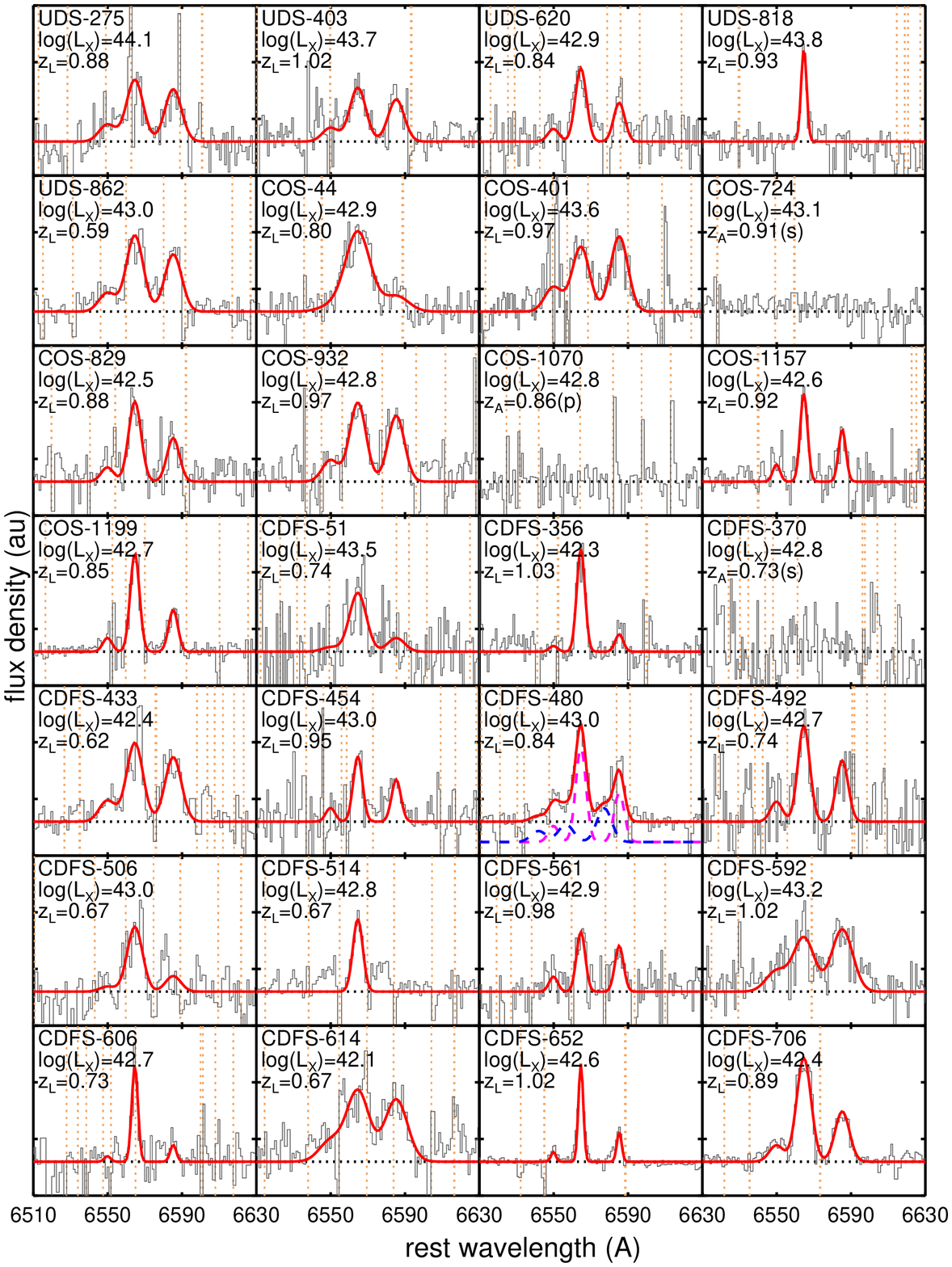,angle=0,width=0.9\textwidth}
}
\caption{Galaxy-integrated spectra, shifted to the
  rest frame, around the H$\alpha$ and [N~{\sc ii}]6548,6583
  emission-lines, in arbitary flux density units, for the
  $z\approx$\,0.6--1.1 targets that do not show a BLR component (i.e.,
  only the Type~2 sources). For descriptions of the different curves see Figure~\ref{fig:haexamples}. The vertical dotted lines indicate the wavelengths of the
  brightest sky lines (\protect\citealt{Rousselot00}). In each panel we identify the target
  name, the logarithm of the X-ray luminosity (in units of erg\,s$^{-1}$) and
  the redshift (see Table~\ref{tab:targets}).} 
\label{fig:specha} 
\end{figure*} 

\begin{figure*} 
\centerline{
\psfig{figure=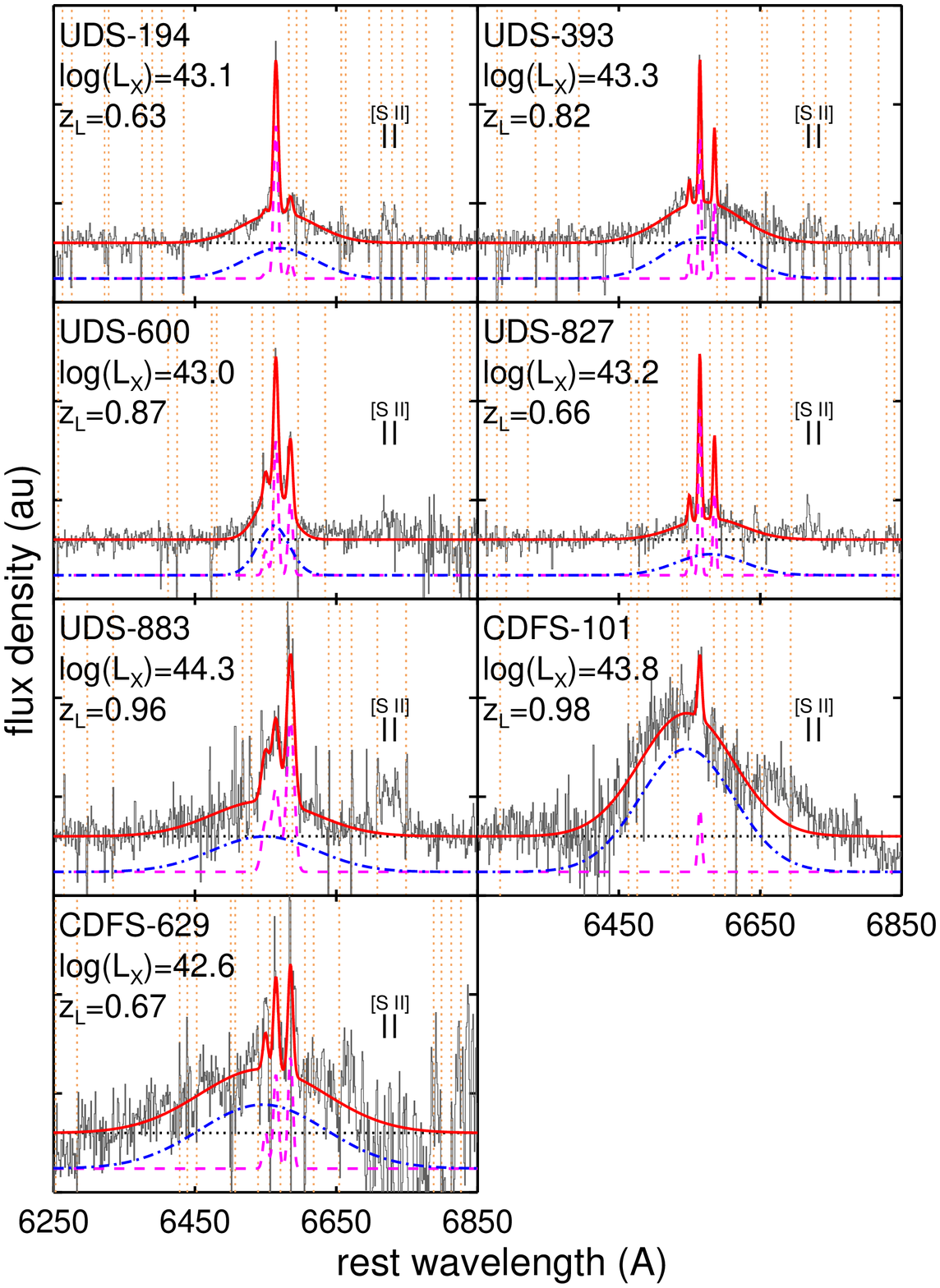,angle=0,width=0.9\textwidth}
}
\caption{Same as Figure~\ref{fig:specha} but for the targets with a
  H$\alpha$ BLR component (i.e., the Type~1 sources). In each panel, the small thick lines show
the wavelengths of the [S~{\sc ii}]6716,6731 emission-line doublet.} 
\label{fig:spechablr} 
\end{figure*} 

\begin{center}

\begin{landscape}
\begin{table}
{\footnotesize
{\centerline {\sc KASH$z$ Target Properties}}
\begin{tabular}{llcccccccccccccccccc}
\hline
Name & $z_{A}$ & Type & $F_{{\rm 0.5-2}}$ & $F_{{\rm 2-10}}$ & Obs. & RL & Inst. & Line & t$_{e}$ & Note &BLR & $z_{L}$ & $S_{A}$ & FW$_{A}$ & $S_{B}$ & FW$_{B}$ & $\Delta v$ & $S_{{\rm [N II]}}$ & $W_{80}$ \\
(1) & (2) & (3) & (4) & (5) & (6) & (7) & (8) & (9) & (10) & (11) & (12) & (13) & (14) & (15) & (16) & (17) & (18) & (19) & (20)\\
\hline
UDS-186   & 1.437 & s & 4.07 & 5.06 & U & \xmark & K & O &  9.0 & C & - & - & 
$<$0.7 & - & - & - & - & - & - \\
UDS-310   & 1.645 & s & 6.96 & 14.15 & U & \xmark & K & O &  5.4 & L & - & 1.645
 & 1.8$\pm$0.7 & 176$\pm$86 & - & - & - & - & 192$\pm$94 \\
UDS-357   & 1.100 & s & 16.16 & 33.18 & U & \xmark & K & O &  5.4 & L & $\beta$
 & 1.098 & 4.5$\pm$1.0 & 779$\pm$154 & - & - & - & - & 848$\pm$168 \\
UDS-574   & 1.533 & s & 4.55 & 4.51 & U & \xmark & K & O &  6.0 & C & - & - & 
$<$0.9 & - & - & - & - & - & - \\
UDS-584   & 1.602 & s & 2.74 & 5.22 & U & \xmark & K & O &  9.0 & L & - & 1.605
 & 4.5$\pm$0.8 & 107$\pm$26 & 25.0$\pm$1.1 & 611$\pm$28 & -52$\pm$13 & - & 
615$\pm$26 \\
UDS-586   & 1.171 & s & 11.97 & 16.01 & U & \xmark & K & O &  6.0 & L & - & 
1.171 & 2.3$\pm$0.3 & 575$\pm$83 & - & - & - & - & 625$\pm$90 \\
UDS-605   & 1.096 & s & 2.77 & 9.76 & O & \xmark & K & O &  6.0 & L & - & 1.097
 & 1.7$\pm$0.7 & 696$\pm$253 & - & - & - & - & 758$\pm$276 \\
UDS-619   & 1.410 & s & 19.18 & 35.01 & U & \xmark & K & O &  6.0 & L & $\beta$
 & 1.401 & 21.2$\pm$4.9 & 242$\pm$29 & 33.8$\pm$4.3 & 666$\pm$68 & -18$\pm$15 & 
- & 562$\pm$33 \\
UDS-631   & 1.192 & s & 3.79 & 10.70 & U & \xmark & K & O &  8.4 & L & - & 1.190
 & 1.6$\pm$0.4 & 318$\pm$95 & - & - & - & - & 346$\pm$104 \\
UDS-640   & 1.092 & s & 6.61 & 10.32 & U & \xmark & K & O &  6.0 & L & - & 1.093
 & 5.4$\pm$0.3 & 523$\pm$36 & - & - & - & - & 570$\pm$39 \\
UDS-643   & 1.087 & s & 3.46 & 5.45 & U & \xmark & K & O &  6.0 & L & $\beta$ & 
1.088 & 2.4$\pm$0.7 & 146$\pm$66 & 2.8$\pm$0.9 & 565$\pm$149 & -189$\pm$121 & -
 & 492$\pm$161 \\
UDS-655   & 1.397 & s & 4.92 & 8.32 & U & \xmark & K & O &  6.0 & L & - & 1.397
 & 1.5$\pm$0.6 & 272$\pm$114 & - & - & - & - & 297$\pm$125 \\
UDS-671   & 1.083 & s & 0.27 & 8.02 & O & \xmark & K & O &  6.0 & L & - & 1.084
 & 0.8$\pm$0.5 & 77$\pm$66 & 3.6$\pm$0.8 & 712$\pm$176 & -191$\pm$90 & - & 
702$\pm$134 \\
UDS-676   & 1.086 & s & 3.41 & 1.91 & U & \xmark & K & O &  9.0 & L & - & 1.086
 & 0.7$\pm$0.3 & 389$\pm$215 & - & - & - & - & 424$\pm$235 \\
UDS-701   & 1.653 & s & 1.45 & 3.35 & U & \xmark & K & O &  9.0 & L & - & 1.649
 & 2.1$\pm$1.1 & 520$\pm$268 & - & - & - & - & 566$\pm$291 \\
UDS-763   & 1.413 & s & 18.88 & 26.59 & U & \xmark & K & O &  9.0 & L & $\beta$
 & 1.412 & 5.4$\pm$3.0 & 261$\pm$78 & 12.3$\pm$3.3 & 705$\pm$142 & -168$\pm$90
 & - & 668$\pm$84 \\
UDS-796   & 1.132 & s & 3.80 & 29.03 & O & \xmark & K & O &  9.0 & C & - & - & 
$<$1.7 & - & - & - & - & - & - \\
UDS-814   & 1.074 & s & 25.01 & 22.10 & U & \xmark & K & O &  9.0 & L & - & 
1.077 & 3.1$\pm$0.5 & 735$\pm$133 & - & - & - & - & 800$\pm$144 \\
UDS-855   & 1.407 & s & 5.19 & 20.10 & O & \cmark & K & O &  8.4 & L & - & 1.407
 & 11.4$\pm$1.1 & 475$\pm$45 & - & - & - & - & 517$\pm$49 \\
COS-20    & 1.156 & s & 1.34 & 7.01 & O & \cmark & K & O &  9.6 & L & - & 1.154
 & 1.7$\pm$0.5 & 324$\pm$130 & 0.9$\pm$0.6 & 380$\pm$214 & 381$\pm$25 & - & 
612$\pm$120 \\
COS-27$^{\star}$ & 1.510 & p & $<$0.73 & 2.59 & O & \xmark & K & O &  5.4 & C & 
- & - & - & - & - & - & - & - & - \\
COS-40    & 1.510 & s & 4.69 & 9.69 & U & \xmark & K & O &  9.0 & L & - & 1.504
 & 2.8$\pm$1.3 & 485$\pm$171 & 2.8$\pm$1.6 & 1277$\pm$375 & 0$\pm$72 & - & 
953$\pm$273 \\
COS-61    & 1.478 & s & 1.45 & 4.41$\dagger$ & ? & \xmark & K & O &  8.4 & L & -
 & 1.478 & 3.5$\pm$0.6 & 225$\pm$45 & - & - & - & - & 245$\pm$49 \\
COS-108   & 1.253 & s & 4.39 & 16.10 & O & \cmark & S & O & 12.0 & L & - & 1.258
 & 25.6$\pm$3.0 & 331$\pm$26 & 10.9$\pm$4.6 & 1360$\pm$409 & -506$\pm$220 & - & 
936$\pm$170 \\
COS-178   & 1.347 & s & 1.47 & 8.92 & O & \xmark & S & O & 12.0 & C & - & - & 
$<$1.2 & - & - & - & - & - & - \\
COS-203   & 1.360 & s & 7.37 & 14.20 & U & \xmark & K & O &  7.2 & L & - & 1.359
 & 5.1$\pm$0.6 & 337$\pm$49 & - & - & - & - & 367$\pm$53 \\
COS-206   & 1.483 & s & 5.89 & 10.40 & U & \xmark & K & O &  7.2 & L & $\beta$
 & 1.480 & 3.8$\pm$2.0 & 630$\pm$323 & - & - & - & - & 685$\pm$352 \\
COS-211   & 1.166 & s & 2.65 & 4.12 & U & \xmark & K & O &  7.2 & L & - & 1.167
 & 2.4$\pm$0.5 & 535$\pm$113 & - & - & - & - & 582$\pm$123 \\
COS-453   & 1.625 & s & 3.01 & 5.70 & U & \cmark & K & O &  9.0 & C & - & - & 
$<$0.5 & - & - & - & - & - & - \\
COS-454   & 1.478 & s & 5.91 & 10.90 & U & \xmark & K & O &  9.0 & L & $\beta$
 & 1.484 & 2.5$\pm$1.3 & 250$\pm$110 & 3.9$\pm$1.9 & 827$\pm$415 & -480$\pm$198
 & - & 922$\pm$280 \\
COS-499   & 1.459 & s & 27.40 & 57.80 & U & \cmark & K & O &  9.0 & L & $\beta$
 & 1.455 & 3.6$\pm$1.2 & 165$\pm$49 & 22.1$\pm$1.7 & 786$\pm$58 & -109$\pm$28 & 
- & 796$\pm$52 \\
COS-649   & 1.369 & s & $<$1.05 & 3.77 & O & \xmark & K & O &  5.4 & L & - & 
1.367 & 2.0$\pm$0.7 & 415$\pm$166 & - & - & - & - & 451$\pm$180 \\
COS-832$^{\star}$ & 1.471 & p & $<$0.86 & 2.71 & O & \cmark & K & O &  8.4 & C
 & - & - & - & - & - & - & - & - & - \\
COS-885$^{\star}$ & 1.594 & p & 0.44 & 1.34$\dagger$ & ? & \xmark & K & O &  9.0
 & C & - & - & - & - & - & - & - & - & - \\
COS-1015  & 1.379 & s & 1.14 & 3.47$\dagger$ & U & \xmark & K & O &  7.2 & L & -
 & 1.377 & 1.4$\pm$0.6 & 281$\pm$160 & - & - & - & - & 306$\pm$174 \\
COS-1518$^{\star}$ & 1.122 & s & $<$0.24 & 2.90 & O & \xmark & K & O &  9.6 & N
 & - & - & - & - & - & - & - & - & - \\
COS-3178  & 1.355 & s & $<$0.59 & 2.54 & O & \xmark & K & O &  7.2 & L & - & 
1.354 & 3.6$\pm$2.3 & 882$\pm$547 & - & - & - & - & 960$\pm$596 \\
COS-3715  & 1.103 & s & 0.36 & 1.11$\dagger$ & ? & \xmark & K & O &  5.4 & C & -
 & - & $<$0.7 & - & - & - & - & - & - \\
XCOS-2028 & 1.592 & s & 33.50 & 75.70 & U & \cmark & S & O & 25.2 & L & $\beta$
 & 1.593 & 17.3$\pm$5.1 & 410$\pm$36 & 16.4$\pm$5.4 & 657$\pm$88 & -332$\pm$76
 & - & 734$\pm$17 \\
XCOS-5627 & 1.337 & s & 7.25 & 15.40 & U & \xmark & S & O & 12.6 & L & $\beta$
 & 1.349 & 0.7$\pm$0.7 & 62$\pm$62 & 9.0$\pm$2.1 & 573$\pm$191 & -5$\pm$14 & -
 & 602$\pm$131 \\
CDFS-25   & 1.374 & s & $<$0.19 & 1.21 & O & \xmark & K & O &  6.0 & L & - & 
1.377 & 3.5$\pm$2.1 & 184$\pm$51 & 9.7$\pm$2.4 & 518$\pm$94 & -183$\pm$64 & - & 
524$\pm$37 \\
\hline
\end{tabular}
}
\caption{Continued over page.}
\end{table}
\end{landscape}
\newpage
\begin{landscape}
\addtocounter{table}{-1}
\begin{table}
{\footnotesize
{\centerline {\sc KASH$z$ Target Properties (continued)}}
\begin{tabular}{llcccccccccccccccccc}
\hline
Name & $z_{A}$ & Type & $F_{{\rm 0.5-2}}$ & $F_{{\rm 2-10}}$ & Obs. & RL & Inst. & Line & t$_{e}$ & Note &BLR & $z_{L}$ & $S_{A}$ & FW$_{A}$ & $S_{B}$ & FW$_{B}$ & $\Delta v$ & $S_{{\rm [N II]}}$ & $W_{80}$ \\
(1) & (2) & (3) & (4) & (5) & (6) & (7) & (8) & (9) & (10) & (11) & (12) & (13) & (14) & (15) & (16) & (17) & (18) & (19) & (20)\\
\hline
CDFS-88   & 1.616 & s & 3.07 & 4.67 & U & \xmark & K & O &  9.0 & C & - & - & 
$<$0.7 & - & - & - & - & - & - \\
CDFS-113$^{\star}$ & 1.608 & i & 0.57 & 0.98 & U & \xmark & K & O &  9.0 & N & -
 & - & - & - & - & - & - & - & - \\
CDFS-166  & 1.605 & s & 8.14 & 12.39 & U & \cmark & K & O &  9.0 & L & - & 1.611
 & 5.7$\pm$0.4 & 371$\pm$25 & - & - & - & - & 404$\pm$27 \\
CDFS-191  & 1.185 & s & $<$0.13 & 4.18 & O & \xmark & K & O &  6.0 & L & - & 
1.184 & 1.8$\pm$0.6 & 297$\pm$125 & - & - & - & - & 323$\pm$136 \\
CDFS-243  & 1.097 & s & 0.71 & 12.29 & O & \xmark & K & O &  6.0 & L & - & 1.096
 & 2.5$\pm$0.8 & 468$\pm$160 & - & - & - & - & 510$\pm$174 \\
CDFS-343  & 1.090 & s & $<$0.08 & 1.46 & O & \xmark & K & O &  9.6 & C & - & -
 & $<$0.4 & - & - & - & - & - & - \\
CDFS-344  & 1.617 & s & 3.64 & 3.89 & U & \cmark & K & O &  7.8 & L & - & 1.613
 & 3.7$\pm$0.9 & 258$\pm$40 & 7.4$\pm$1.4 & 1089$\pm$149 & -270$\pm$83 & - & 
977$\pm$112 \\
CDFS-518  & 1.603 & s & 9.74 & 13.30 & U & \cmark & K & O &  9.6 & L & - & 1.609
 & 8.3$\pm$1.0 & 850$\pm$88 & - & - & - & - & 926$\pm$96 \\
CDFS-549$^{\star}$ & 1.553 & s & 1.36 & 3.25 & U & \xmark & K & O & 11.4 & N & -
 & - & - & - & - & - & - & - & - \\
CDFS-619  & 1.178 & s & 0.16 & 2.17 & O & \xmark & K & O &  7.8 & L & - & 1.179
 & 1.6$\pm$0.6 & 419$\pm$147 & - & - & - & - & 457$\pm$160 \\
CDFS-720  & 1.609 & s & 10.93 & 18.44 & U & \xmark & K & O &  9.0 & L & - & 
1.610 & 2.2$\pm$0.5 & 742$\pm$173 & - & - & - & - & 808$\pm$189 \\
SSA22-39  & 1.397 & s & 7.19 & 9.08 & U & ? & S & O &  3.0 & L & - & 1.400 & 
8.2$\pm$1.8 & 605$\pm$143 & - & - & - & - & 658$\pm$156 \\
SSA22-141 & 1.370 & s & 20.53 & 27.38 & U & ? & S & O &  2.4 & L & - & 1.367 & 
5.4$\pm$2.2 & 541$\pm$209 & - & - & - & - & 589$\pm$227 \\
UDS-194   & 0.627 & s & 5.80 & 10.13 & U & \xmark & K & H &  9.0 & L & $\alpha$
 & 0.628 & 8.1$\pm$0.9 & 328$\pm$42 & 25.2$\pm$2.3 & 5385$\pm$604 & 258$\pm$190
 & 0.93 & 357$\pm$46 \\
UDS-275   & 0.883 & s & 7.27 & 52.96 & O & \xmark & K & H &  5.4 & L & - & 0.882
 & 4.5$\pm$0.9 & 476$\pm$87 & - & - & - & 3.85 & 518$\pm$95 \\
UDS-393   & 0.822 & s & 9.70 & 9.38 & U & \xmark & K & H &  9.0 & L & $\alpha$
 & 0.822 & 6.0$\pm$0.5 & 183$\pm$22 & 49.1$\pm$3.4 & 5811$\pm$457 & 176$\pm$164
 & 3.18 & 199$\pm$24 \\
UDS-403   & 1.021 & s & 2.30 & 13.81 & O & \xmark & K & H &  9.0 & L & - & 1.021
 & 3.7$\pm$0.9 & 461$\pm$118 & - & - & - & 2.89 & 502$\pm$129 \\
UDS-600   & 0.873 & s & 4.41 & 4.34 & U & \xmark & K & H &  6.0 & L & $\alpha$
 & 0.873 & 11.6$\pm$1.3 & 343$\pm$36 & 29.2$\pm$3.1 & 2401$\pm$216 & 
-158$\pm$108 & 6.39 & 373$\pm$39 \\
UDS-620   & 0.842 & s & 1.08 & 3.44 & O & \xmark & K & H &  9.0 & L & - & 0.843
 & 2.3$\pm$0.4 & 311$\pm$75 & - & - & - & 1.21 & 338$\pm$81 \\
UDS-818   & 0.928 & s & 14.55 & 19.28 & U & \xmark & K & H &  8.4 & L & - & 
0.928 & 5.5$\pm$0.6 & 140$\pm$26 & - & - & - & $<$0.66 & 153$\pm$29 \\
UDS-827   & 0.658 & s & 4.65 & 12.22 & U & \xmark & K & H &  8.4 & L & $\alpha$
 & 0.657 & 8.0$\pm$0.4 & 201$\pm$15 & 22.7$\pm$2.1 & 5129$\pm$548 & 682$\pm$196
 & 4.02 & 219$\pm$17 \\
UDS-862   & 0.589 & s & 1.35 & 8.50 & O & \xmark & K & H &  8.4 & L & - & 0.589
 & 3.8$\pm$0.5 & 471$\pm$64 & - & - & - & 2.88 & 512$\pm$70 \\
UDS-883   & 0.961 & s & 24.39 & 56.58 & U & \xmark & K & H &  8.4 & L & $\alpha$
 & 0.961 & 5.2$\pm$0.4 & 574$\pm$29 & 30.5$\pm$2.9 & 7970$\pm$795 & -748$\pm$271
 & 9.50 & 625$\pm$32 \\
COS-44    & 1.513 & p & 1.23 & 3.56 & U & \xmark & K & H &  9.0 & L & - & 0.801
 & 4.9$\pm$0.4 & 709$\pm$87 & - & - & - & 0.95 & 771$\pm$95 \\
COS-401   & 0.969 & s & $<$1.14 & 13.30 & O & \cmark & S & H &  2.4 & L & - & 
0.971 & 7.6$\pm$0.9 & 495$\pm$53 & - & - & - & 8.80 & 539$\pm$58 \\
COS-724   & 0.906 & s & $<$0.51 & 4.04 & O & \xmark & K & H &  6.0 & C & - & -
 & $<$1.2 & - & - & - & - & - & - \\
COS-829   & 0.885 & s & 0.39 & 1.17$\dagger$ & ? & \xmark & K & H &  8.4 & L & -
 & 0.885 & 6.4$\pm$0.8 & 353$\pm$53 & - & - & - & 3.50 & 384$\pm$57 \\
COS-932   & 0.975 & s & 0.71 & 2.17$\dagger$ & U & \xmark & K & H &  8.4 & L & -
 & 0.974 & 6.5$\pm$0.6 & 469$\pm$49 & - & - & - & 5.46 & 511$\pm$53 \\
COS-1070$^{\star}$ & 0.858 & p & 1.06 & 2.30 & U & \xmark & K & H &  8.4 & C & -
 & - & - & - & - & - & - & - & - \\
COS-1157  & 0.915 & s & 0.49 & 1.50$\dagger$ & ? & \xmark & K & H &  9.6 & L & -
 & 0.925 & 5.2$\pm$0.6 & 197$\pm$29 & - & - & - & 3.07 & 215$\pm$31 \\
COS-1199  & 0.771 & p & 0.67 & 2.10 & O & \xmark & K & H &  9.0 & L & - & 0.850
 & 9.5$\pm$0.7 & 255$\pm$23 & - & - & - & 4.05 & 277$\pm$25 \\
CDFS-51   & 0.737 & s & 1.81 & 18.33 & O & \xmark & S & H &  2.7 & L & - & 0.737
 & 11.9$\pm$3.5 & 481$\pm$211 & - & - & - & $<$3.79 & 523$\pm$230 \\
CDFS-101  & 0.977 & s & 11.57 & 20.66 & U & \xmark & K & H &  9.0 & L & $\alpha$
 & 0.978 & 2.9$\pm$0.5 & 290$\pm$68 & 133.5$\pm$3.7 & 7150$\pm$202 & -843$\pm$77
 & $<$2.78 & 315$\pm$74 \\
CDFS-356  & 1.034 & s & 0.29 & 0.57 & U & \xmark & K & H &  6.0 & L & - & 1.034
 & 3.3$\pm$0.2 & 223$\pm$17 & - & - & - & 0.55 & 243$\pm$19 \\
CDFS-370  & 0.734 & s & 1.07 & 3.38 & O & \xmark & S & H &  2.7 & C & - & - & 
$<$3.8 & - & - & - & - & - & - \\
CDFS-433  & 0.617 & s & 1.04 & 2.03 & U & \xmark & K & H &  9.6 & L & - & 0.620
 & 3.7$\pm$0.4 & 461$\pm$52 & - & - & - & 3.03 & 501$\pm$57 \\
CDFS-454  & 0.952 & i & 1.42 & 3.17 & U & \xmark & K & H & 11.4 & L & - & 0.953
 & 2.0$\pm$0.5 & 257$\pm$82 & - & - & - & 1.30 & 279$\pm$89 \\
CDFS-480  & 0.839 & s & 2.48 & 4.08 & U & \xmark & K & H &  6.0 & L & - & 0.841
 & 7.3$\pm$2.1 & 286$\pm$41 & 2.0$\pm$2.1 & 427$\pm$280 & -367$\pm$169 & 4.80 & 
521$\pm$105 \\
CDFS-492  & 0.735 & s & 0.13 & 2.55 & O & \xmark & S & H &  5.4 & L & - & 0.735
 & 11.2$\pm$0.6 & 334$\pm$22 & - & - & - & 7.13 & 363$\pm$24 \\
CDFS-506  & 0.665 & s & 4.30 & 7.35 & U & \xmark & K & H &  7.8 & L & - & 0.666
 & 1.8$\pm$0.4 & 430$\pm$109 & - & - & - & $<$0.46 & 468$\pm$119 \\
\hline
\end{tabular}
}
\caption{Continued over page.}
\end{table}
\end{landscape}
\newpage
\begin{landscape}
\addtocounter{table}{-1}
\begin{table}
{\footnotesize
{\centerline {\sc KASH$z$ Target Properties (continued)}}
\begin{tabular}{llcccccccccccccccccc}
\hline
Name & $z_{A}$ & Type & $F_{{\rm 0.5-2}}$ & $F_{{\rm 2-10}}$ & Obs. & RL & Inst. & Line & t$_{e}$ & Note &BLR & $z_{L}$ & $S_{A}$ & FW$_{A}$ & $S_{B}$ & FW$_{B}$ & $\Delta v$ & $S_{{\rm [N II]}}$ & $W_{80}$ \\
(1) & (2) & (3) & (4) & (5) & (6) & (7) & (8) & (9) & (10) & (11) & (12) & (13) & (14) & (15) & (16) & (17) & (18) & (19) & (20)\\
\hline
CDFS-514  & 0.664 & s & 3.29 & 4.49 & U & \xmark & K & H &  7.8 & L & - & 0.667
 & 4.2$\pm$0.8 & 265$\pm$81 & - & - & - & $<$0.99 & 289$\pm$88 \\
CDFS-561  & 0.798 & s & 0.08 & 2.69 & O & \xmark & K & H &  7.8 & L & - & 0.980
 & 1.4$\pm$0.4 & 239$\pm$91 & - & - & - & 1.11 & 261$\pm$99 \\
CDFS-592  & 1.016 & s & 0.69 & 4.03 & O & \xmark & K & H &  9.0 & L & - & 1.015
 & 5.3$\pm$0.9 & 601$\pm$89 & - & - & - & 6.13 & 655$\pm$97 \\
CDFS-606  & 0.733 & s & 0.34 & 2.58 & O & \xmark & K & H &  7.8 & L & - & 0.733
 & 1.7$\pm$0.2 & 151$\pm$17 & - & - & - & 0.29 & 164$\pm$19 \\
CDFS-614  & 0.668 & s & 0.38 & 0.86 & U & \xmark & K & H &  9.0 & L & - & 0.667
 & 4.7$\pm$0.5 & 642$\pm$84 & - & - & - & 4.09 & 699$\pm$92 \\
CDFS-629  & 0.667 & s & 0.83 & 2.60 & O & \xmark & K & H &  9.0 & L & $\alpha$
 & 0.667 & 1.7$\pm$0.5 & 360$\pm$89 & 30.1$\pm$6.3 & 9647$\pm$1706 & 
-846$\pm$549 & 2.04 & 392$\pm$97 \\
CDFS-652  & 1.020 & s & 0.66 & 1.18 & U & \xmark & K & H &  7.8 & L & - & 1.022
 & 8.0$\pm$0.7 & 115$\pm$21 & - & - & - & 2.40 & 125$\pm$22 \\
CDFS-706  & 0.891 & s & $<$0.15 & 1.02 & O & \xmark & K & H &  9.0 & L & - & 
0.890 & 13.7$\pm$0.4 & 409$\pm$16 & - & - & - & 6.63 & 445$\pm$18 \\
\hline
\hline
\label{Tab:Targets}
\end{tabular}
}
\caption{\label{tab:targets}
Properties for the KASH$z$ targets observed so far. Notes:
(1) Source name (`field' -- `X-ray ID'), those targets followed by a
$^{\star}$ are excluded from the analyses presented in this paper (see
Section~\ref{sec:detectionrates}); (2) archival redshift; (3) archival
redshift type (i.e., photometric [p], secure spectroscopic redshift
[s] or insecure spectroscopic redshift [i]); (4) and (5) soft-band
(0.5--2\,keV) and hard-band (2--10\,keV)
X-ray fluxes ($\times$10$^{-15}$\,erg\,s$^{-1}$\,cm\,$^{-2}$), where
the hard-band values followed by a $\dagger$ were
estimated from the soft-band fluxes (see Section~\ref{sec:sample});
(6) flag to indicate X-ray obscured candidates (O) and unobscured
candidates (U), those without sufficient constraints are labelled with
``?''; (7) flag to indicate the ``radio luminous'' targets with
$L_{{\rm 1.4GHz}}>10^{24}$\,W\,Hz$^{-1}$ (Section~\ref{sec:radiodata}); (8) instrument used for the observations (K$=$KMOS and S$=$SINFONI); (9) primary targeted
emission line (O$=$[O~{\sc iii}] and H$=$H$\alpha$); (10) on-source
exposure time (kilo-seconds); (11) note on data (L$=$line detected;
C$=$ continuum only detected; N$=$no detection); (12) note for the
identification of a BLR component ($\alpha$ for H$\alpha$ and $\beta$ for
H$\beta$); (13) redshift derived from the narrowest Gaussian component of the emission-line
profile fit; (14) and (15) Flux ($\times10^{-17}$\,erg\,s$^{-1}$\,cm$^{-2}$) and FWHM
(km\,s$^{-1}$) of the narrower Gaussian component; (16) and (17) Flux ($\times10^{-17}$\,erg\,s$^{-1}$\,cm$^{-2}$) and FWHM  of
the broader Gaussian component (for sources with ``$\alpha$'' in
column 10, this is the BLR component); (18) velocity offset between
the two Gaussian components (km\,s$^{-1}$); (19) flux of [N~{\sc
  ii}]$\lambda$6583 ($\times10^{-17}$\,erg\,s$^{-1}$\,cm$^{-2}$) where
applicable; (20) overall emission-line width (km\,s$^{-1}$). All of
the quoted uncertainties are the random errors on the fits due to the
noise in the spectra; however, we note that throughout this work we
add an extra 30\% systematic error to the emission-line fluxes to
account for the uncertainty in the flux calibration (see Section~\ref{sec:linefitting}). 
}
\end{table}
\end{landscape}

\end{center}

\end{document}